# Tunable Energy Level Alignment in the Multilayers of Carboxylic Acids on Silver

Veronika Stará,[1] Pavel Procházka,[1] Jakub Planer,[1] Azin Shahsavar,[1] Anton O. Makoveev,[1] Tomáš Skála,[2] Matthias Blatnik,[1] Jan Čechal[1,3*]

[1] CEITEC - Central European Institute of Technology, Brno University of Technology, Purkyňova 123, 612 00 Brno, Czech Republic.

[2] Department of Surface and Plasma Science, Faculty of Mathematics and Physics, Charles University, V Holešovičkách 2, 18000 Prague 8, Czech Republic

[3] Institute of Physical Engineering, Brno University of Technology, Technická 2896/2, 616 69 Brno, Czech Republic.

AUTHOR INFORMATION

Corresponding Author

* E-mail: cechal@fme.vutbr.cz (J. Č.)

ABSTRACT

The precise energy level alignment between a metal electrode and an organic semiconductor is required to reduce contact resistance and enhance the efficiency of organic-semiconductor-based devices. Here, we introduce monolayer thick charge injection layers (CILs) based on aromatic carboxylic acids that can induce an energy level shift in the subsequent layers by up to 0.8 eV. By a gradual chemical transformation of the as-deposited molecules, we achieve a highly tunable energy level shift in the range of 0.5 eV. We reveal that the work function and energy-level positions in the CIL increase linearly with the density of induced dipoles. The energy level position of the subsequent layers follows the changes in the CIL. Our results thus connect the energy alignment quantities, and the high tunability would allow precise tuning of the active layers deposited on the CIL, which marks a path towards efficient charge injection layers on metal electrodes.

TEXT

## 1. Introduction

Organic electronics gained a significant position in the illumination and display technologies [1–3] with other areas employing organic semiconductors (OSC), i.e., organic thin-film transistors [4] and organic photovoltaics [5] being on edge toward large-scale industrial application. To utilize OSC as active components in fast switching electrical devices, forming a good electrical contact is essential. [6–9] The interface between the OSC active layer and the metallic contacts defines the charge injection efficiency of the device, e.g., the alignment of the molecular orbital levels of the organic layer with the vacuum and Fermi levels of the metal electrodes determines the electron- and hole-injection, and a considerable contact resistance arises from their misalignment [7,10–12]. The high contact resistance limits the operation frequency and restricts high current devices such as organic field-effect transistors [7,8].

The introduction of ordered dipolar layers at the metal-OSC interface can effectively tune the electrode work function (WF) and the interfacial energy level alignment (ELA) and, correspondingly, the charge-carrier injection barriers into adsorbate layers. [13]. Passivation of the metal electrode by a thin insulating layer can establish the ELA through integer charge transfer via the tunneling barrier [14,15]. However, the tunneling contact is associated with considerable contact resistance and energy losses despite the precise alignment [16]. Hence, the direct contact of molecules with the electrode surface is preferable. In this case, the deposited molecules can form interfacial dipolar layers and act as so-called charge injection layers (CILs) that reduce the energy level misalignment, thus increasing the efficiency of OSC-based devices [17,18]. The dipoles within CILs can be either intrinsic to the adsorbed molecules or form due to the chemical interaction of the adsorbed molecules with the substrate, molecule–substrate charge transfer, or changing the molecular conformation (e.g., its bending) and orientation with respect to the surface [13].

Extensive surface science studies provided a fundamental understanding of the role of dipolar layers in the changes of WF and ELA, as demonstrated for many molecular systems [10,13,19,20]. However, only several molecular systems provided the possibility of ELA tuning, e.g., by pre-coverage [21], bicomponent blending [22], or electric field tuning in disordered layers [23]. Recently, a high tunability of the electrode WF was demonstrated by changing the composition at the dielectric–metal interface of the epitaxial MgO layer on the Ag surface [24], but the tunneling barrier still features a considerable contact resistance. A precise tunability by a homogeneous single-component molecular layer would provide a significant advantage for the fabrication of low contact resistance interfaces. In this respect, a recent theoretical study introduced the possibility of tuning the work function in the range of 1 eV by the change between electron donor and acceptor character by on-surface chemical interconversion of deposited molecules [25]. Following this concept, we employ 4,4'-biphenyl dicarboxylic acid (BDA, Figure 1a) to prepare a range of single-layer molecular CILs with a tunable density of interface dipoles that are formed by a chemical transformation of the BDA molecules in direct contact with the metal substrate. The formed electrical dipoles induce energy shifts of molecular layers by up to 0.8 eV. We have established a linear relationship between the work function (WF) change and the electrostatic shift in the C 1s position within the first molecular layer (referred to as CIL) and the second molecular layer. Finally, we present a continuous tuning of WF and ELA by a gradual deprotonation of BDA.

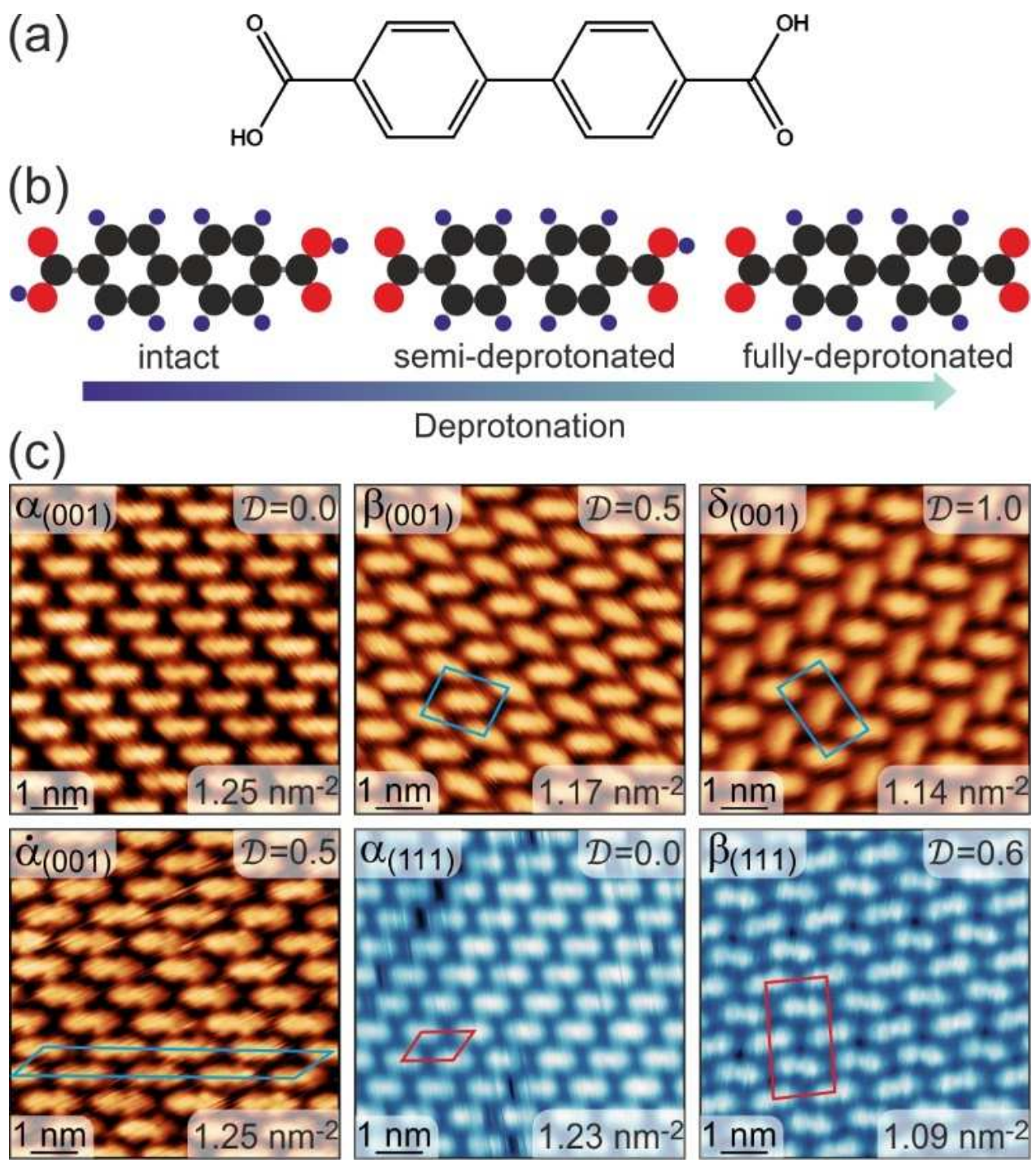

**Figure 1**: The molecule and CILs employed in this study. (a) Chemical structure of 4,4'-biphenyl dicarboxylic acid (BDA). (b) Partially- and fully deprotonated BDA has a similar structure but largely differing properties compared to intact BDA. (c) STM images of distinct CILs considered in this study. The degree of the deprotonation $\mathcal{D}$ is given for each phase. The quadrangles mark the BDA unit cells (note that there is no unit cell for the $\alpha_{(001)}$ phase). The density of BDA molecules $n_{\mathrm{BDA}}$ per nm$^{-2}$ of every CIL is given in the bottom right corner.

## 2. Methods

### 2.1 Experimental details

Both the CIL and the subsequent layers were deposited under ultra-high vacuum UHV conditions (base pressure below $2 \cdot 10^{-10}$ mbar) on the surfaces of Ag(001) and Ag(111) single crystals first cleaned by repeated cycles of $Ar^+$ sputtering and annealing at 520 °C. All analysis was performed under UHV conditions without exposing the sample to the ambient environment. We have employed synchrotron radiation photoelectron spectroscopy [26] as the primary tool to determine the electronic properties of the CILs; the measurements were performed at the Materials Science Beamline at Elettra synchrotron light source in Trieste. Overall, the experimental details are the same as reported previously [27]. In brief, we used 40.8, 420, 510, and 670 eV excitation energies to measure the valence band, C 1s, Ag $3d_{5/2}$, and O 1s spectra, respectively. Detailed spectra were acquired in a medium area lens mode using 10 eV pass energy integrating 3 (Ag 3d, C 1s) or 25 (O 1s) sweeps with 0.1 s dwell time and 0.05 eV energy step. The total resolution was in the range of 300–550 meV. The peak positions were corrected with respect to the measured Fermi edge of the Ag substrate, and the intensity was normalized to the photo-current of a gold mesh placed in the beamline. In addition, we have applied a 7 V bias to the sample during secondary electron cut-off (SEC) measurements, and, subsequently, we have shifted the spectra according to the measured Fermi level position, setting it to 0.00 eV.

The synchrotron measurements were combined with laboratory measurements in which we developed methodologies to prepare the CILs by UHV deposition and subsequent thermal annealing [27–30]. We have employed low-energy electron microscopy (LEEM) to assess the coverage, low-energy electron diffraction (LEED), and scanning tunneling microscopy (STM) to determine the structure, and X-ray photoelectron spectroscopy (XPS) to measure the chemical composition and electronic properties of the samples; the exact experimental parameters were detailed previously [27–30]. X-ray-induced deprotonation was followed in real-time by XPS on the samples kept at 25 °C. A non-monochromatized Mg Kα X-ray source operating at 300 W power was positioned ~25 mm from the sample surface, which corresponds to ~$8\times10^{10}$ photons/s according to the manufacturer.

## 2.2 Theoretical approaches

All the calculations were based on the Density Functional Theory (DFT) and performed using the Vienna ab initio Simulation Package (VASP). [31] The projector-augmented wave (PAW) [32] method was employed for treating core electrons. For silver, oxygen, hydrogen, and carbon 11 valence electrons ($5s^1 4d^{10}$), 6 valence electrons ($2s^2 2p^4$), 1 valence electron ($1s^1$) and 4 valence electrons ($2s^2 2p^2$), respectively, were expanded in a plane-wave basis set with an energy cut-off set to 450 eV. The Brillouin zone was sampled with a Γ-centered Monkhorst-Pack grid, [33] using more than 17 *k*-points per Å$^{-1}$ for all the structural relaxations and 60 *k*-points per Å$^{-1}$ for the subsequent electronic structure calculations. The ionic relaxations were stopped when all the residual forces became smaller than $4\times10^{-2}$ eV/Å. In all the calculations, total energy, potential, and forces were corrected for finite-size errors. All the slabs consisted of eight layers with lateral cell dimensions corresponding to the optimized silver bulk lattice constants. For the $\alpha_{(001)}$, $\dot{\alpha}_{(001)}$ and $\delta_{(001)}$ phases, the BDA molecular layers were adapted to the silver (001) supercell with matrix notations $\begin{pmatrix} 3 & 2 \\ -2 & 2 \end{pmatrix}$ for the α and $\dot{\alpha}$ phases and $\begin{pmatrix} 3 & 5 \\ -3 & 2 \end{pmatrix}$ for the δ phase. All slabs were separated by a vacuum gap of 18 Å. We used non-local optB88-vdW DFT functional [34] to describe the exchange-correlation energy. Benchmark calculations using PBE-D3 [35] and optB86b-vdW [36] functionals showed no qualitative differences in the evaluated dipole moments.

## 3. Results and Discussion

### 3.1 Linear relationship between work function and core-level shifts in the CIL

#### 3.1.1 BDA molecular phases on silver substrates

The BDA molecule features two phenyl rings that impose its flat-lying geometry on metal surfaces and two carboxyl ($-\mathrm{COOH}$) end-groups that mediate intermolecular hydrogen binding and enable the formation of extended supramolecular assemblies. BDA molecules in direct contact with the silver substrate chemically transform when annealed to elevated temperatures – their carboxyl groups deprotonate, i.e., lose hydrogen (Figure 1b) [37]. Hence, we obtain three chemically distinct molecular species with similar structures. After removing the hydrogen, the resulting carboxylate group possesses a partial negative charge, which leads to a formation of an interfacial dipole and the rearrangement of molecules into a structurally distinct phase. Through the gradual deprotonation, we can prepare a range of molecular phases with a distinct density of dipoles $n_{\mathrm{Dip}}$. In this work, we have considered six distinct stable BDA molecular phases as CILs: the α, $\dot{\alpha}$, β, and δ [27,28,30] on Ag(001) and the α and β phases on Ag(111) [29]; the STM images of these phases are given in Figure 1c. In the following, we will designate the full monolayers of these phases as CILs and denote them by the Greek letter standing for the molecular phase with the orientation of the surface plane in the index, e.g., $\beta_{(001)}$ stands for the β phase on the Ag(001) surface.

These phases differ in the degree of deprotonation $\mathcal{D}$ of carboxyl groups of BDA molecules. We define degree of deprotonation as fraction of deprotonated to the total number of carboxyl groups. The values of $\mathcal{D}$ determined by fitting the O 1s XPS spectra are given along the particular STM images in Figure 1c for all the considered phases. The measured spectra of the $\alpha_{(001)}$ and $\dot{\alpha}_{(001)}$ CILs are given in Figure 2. The O 1s spectrum of the $\alpha_{(001)}$ CIL in Figure 2c shows two peaks: the one at binding energy (BE) of 534.05 eV is associated with hydroxyl ($\mathrm{C-\underline{O}H}$) and the one at 532.65 eV with carbonyl ($\mathrm{C=\underline{O}}$) oxygens of the carboxyl groups. Upon thermal annealing of the $\alpha_{(001)}$ CIL to 70 °C, half of the carboxyl groups deprotonate and the $\dot{\alpha}_{(001)}$ CIL is formed [27]. The O 1s spectrum significantly changes: there is one dominant component at 530.9 eV associated with carboxylate oxygens ($\mathrm{-C\underline{OO}^-}$) and two peaks at 532.8 and 531.45 of the two oxygen atoms of the carboxyl group bound to carboxylate (dark red peaks in Figure 2d) [27,28]. The spectra measured for the other CILs are provided in Supplemental Material, Sections 1 and 2. From the fitted spectra, $\mathcal{D}$ is determined as the intensity ratio of the peaks associated with the deprotonated carboxyl groups and the total intensity of the O 1s peak.

The deprotonated carboxyl groups are associated with a dipole moment: the density of dipoles $n_{\mathrm{Dip}}$ is therefore increasing with $\mathcal{D}$. For each molecular phase the $n_{\mathrm{Dip}}$ is obtained as $n_{\mathrm{Dip}} = 2\mathcal{D}n_{\mathrm{BDA}}$, where $n_{\mathrm{BDA}}$ is the surface density of BDA molecules in the particular phase; the factor 2 reflects two carboxyl groups per BDA. The densities $n_{\mathrm{BDA}}$ were determined from the unit cells established by the local congruence method from low-energy electron diffractograms [27–30]; the $n_{\mathrm{BDA}}$ are provided in the STM images given in Figure 1c. For the considered CILs, the $n_{\mathrm{Dip}}$ ranges from 0 ($\alpha_{(001)}$, $\alpha_{(111)}$), via 1.17 ($\beta_{(001)}$), 1.25 ($\dot{\alpha}_{(001)}$), 1.30 ($\beta_{(111)}$) to 2.28 dipoles per nm$^2$ ($\delta_{(001)}$).

The C 1s spectrum of the $\alpha_{(001)}$ CIL (Figure 2a) shows the main component at 285.25 eV; it can be associated with the C atoms of the phenyl rings. The carboxylic C 1s component is shifted by 4.4 eV to higher BE; the remaining features are associated with an extensive shake-up satellite structure [38,39]. The presence of the extensive shake-up satellite structure complicates the precise fitting of carboxyl-related components, as discussed previously for BDA on the Ag(111) substrate [40]. Therefore, we did not consider the position of carboxyl-group-related components to discuss CIL properties. As carbon atoms do not chemically interact with the substrate (see below), we have employed the C 1s level position to describe electrostatic core-level shifts and ELA of the second BDA layer.

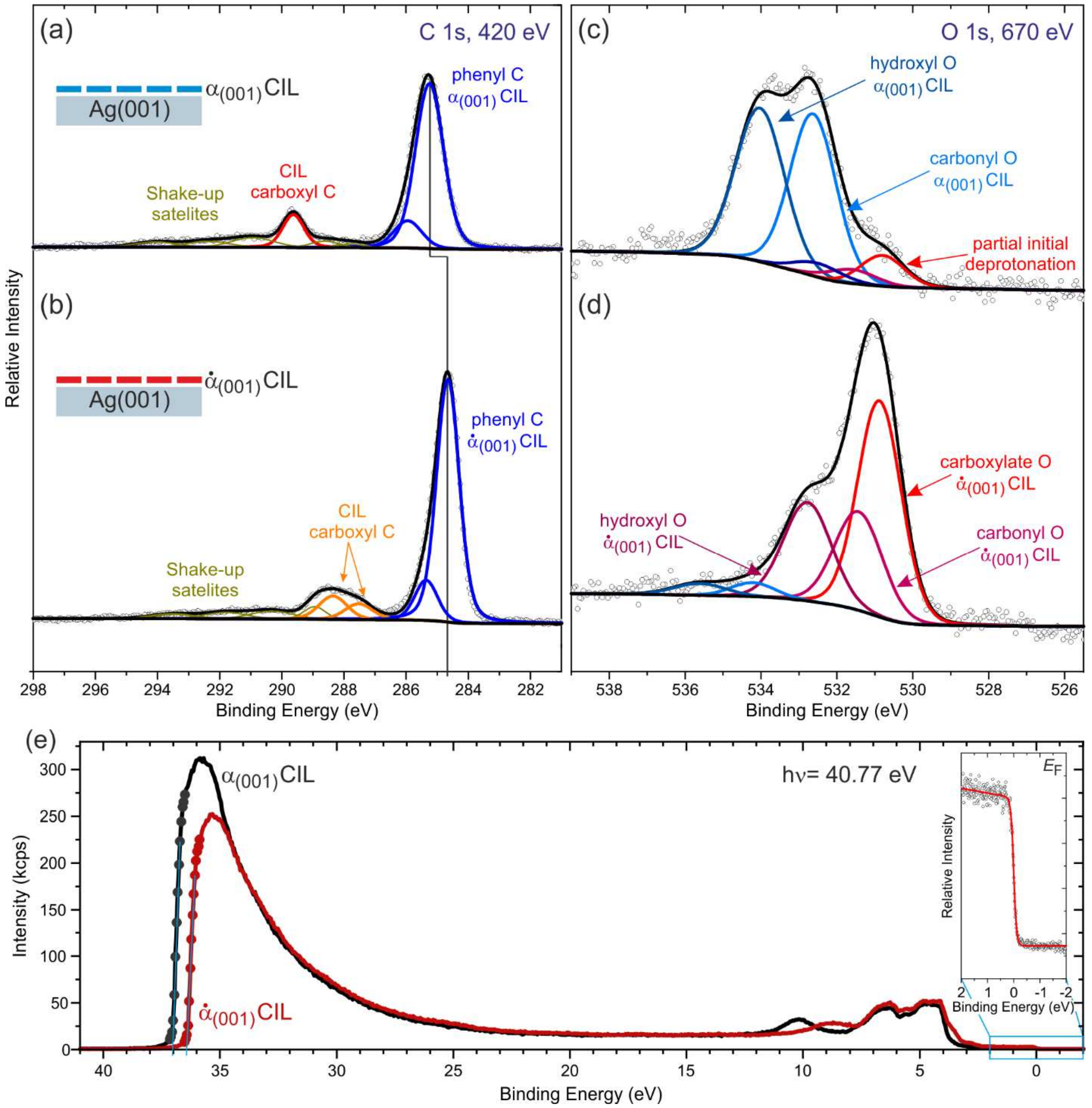

**Figure 2:** Photoelectron spectroscopy analysis of the BDA CILs. (a, b) C 1s spectra measured by synchrotron radiation with an energy of 420 eV from samples comprising the full layer of BDA deposited on Ag(001) substrates, that is $\alpha_{(001)}$ and $\dot{\alpha}_{(001)}$ CILs. The peak components are marked by the associated chemical states; the full description and peak parameters for all CILs are given in Supplemental Material, Sections 1 and 2. The vertical lines mark the shift of the peaks discussed

in the text. (c, d) O 1s spectra (670 eV) of the $\alpha_{(001)}$ and $\dot{\alpha}_{(001)}$ CILs. (e) Wide spectra and detail of the Fermi edge measured on samples with the $\alpha_{(001)}$ and $\dot{\alpha}_{(001)}$ CILs used for the determination of the WF. The full data for all CILs are given in the Supplemental Material, Section 3.

### 3.1.2 Change of work function during the thermal-induced transformation

Upon the deposition of the molecular layer and during its transformation, the sample WF significantly changes. We have determined the WF for all considered CILs and bare substrates employing a standard procedure of measuring a secondary electron cut-off (SEC) and Fermi levels [10,41,42], as demonstrated in Figure 2e for the $\alpha_{(001)}$ and $\dot{\alpha}_{(001)}$ CILs; all the measured SEC are given in Figure S6 in the Supplemental Material, Section 3. For the as-deposited molecular layers, the WF significantly decreases, but with the temperature-induced deprotonation of the carboxyl groups (Figure 3a), the WF increases (Figure 3b). The WF increases linearly with $\mathcal{D}$ but the slope changes when $\mathcal{D}$=0.5 is reached. To get a deeper understanding of these changes we have performed DFT calculations for the $\alpha_{(001)}$, $\dot{\alpha}_{(001)}$ and $\delta_{(001)}$ phases.

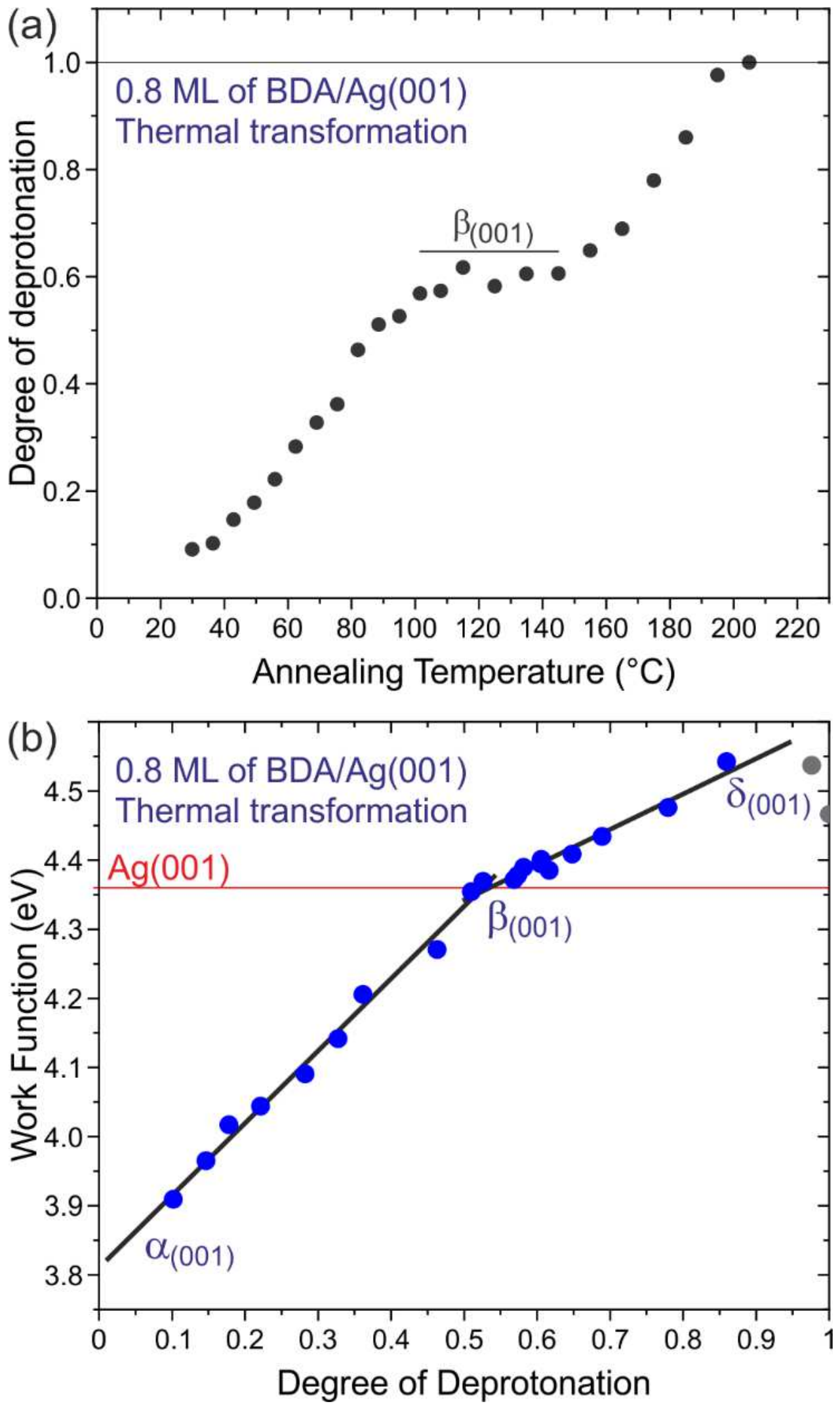

**Figure 3**: (a) Degree of deprotonation of the carboxyl groups obtained during gradual annealing of the sample with 0.8 ML of BDA on the Ag(001) up to 200 °C. After each annealing step, the heating was turned off, and the spectra were measured. The black line indicates the temperatures at which the degree of the deprotonation only slightly changes with temperature; we associate this region with the $\beta_{(001)}$ phase. (b) The change of work function as a function of the degree of deprotonation. The red horizontal line marks the measured substrate work function (4.36 eV). The two last points are given in grey as there is a loss of BDA molecules due to desorption.

### 3.1.3 DFT calculations of a dipole moment in the $\alpha_{(001)}$, $\dot{\alpha}_{(001)}$ and $\delta_{(001)}$ phases

The dipole moment for each phase was calculated for the $\alpha_{(001)}$, $\dot{\alpha}_{(001)}$ and $\delta_{(001)}$ phases (Figures 4a–c) as the sum of interface ($\mu_{\mathrm{int}}$) and intramolecular ($\mu_{\mathrm{mol}}$) dipole moments. The interface dipole moments contain a contribution from the push-back effect and a contribution from the interface dipoles induced by charge transfer. The $\mu_{\mathrm{mol}}$ term contains a dipole moment localized within the molecular layer. It is induced by the bending of BDA molecules (see Figures 4g, h, and i) due to a strong interaction of the deprotonated carboxylic groups with the silver substrate. Absolute values of the interface dipole moments were calculated from the electron density difference ($\Delta\rho$) between the silver substrate and the molecular layer, defined as:

$$\Delta\rho = \rho_{\mathrm{BDA+Ag}} - \left(\rho_{\mathrm{BDA}} + \rho_{\mathrm{Ag}}\right), \qquad (1)$$

where $\rho_{\mathrm{BDA+Ag}}$ is the total electron density of the combined system, $\rho_{\mathrm{Ag}}$ marks the electron density of the silver substrate without the molecular layer, and $\rho_{\mathrm{BDA}}$ is the electron density of a free-standing BDA monolayer. The expression for the interface dipole moments reads:

$$\mu_{\mathrm{int}} = \iiint z \cdot \Delta\rho(x,y,z)\mathrm{d}x\mathrm{d}y\mathrm{d}z. \qquad (2)$$

The intramolecular part of the dipole moment was calculated from the total charge density of the molecular layer, including the ionic contribution:

$$\mu_{\mathrm{mol}} = \iiint z \cdot \rho_{\mathrm{BDA}}\mathrm{d}x\mathrm{d}y\mathrm{d}z - \sum_i^N Z_i \cdot q_i, \qquad (3)$$

where $N$, $Z_i$ and $q_i$ denote the number of ions in the molecular layer, their positions, and charges, respectively. The resulting dipole moment, i.e., $\mu_{\mathrm{tot}} = \mu_{\mathrm{int}} + \mu_{\mathrm{mol}}$ is in perfect agreement (less than 0.06 D in absolute values) with its evaluation from the work function change [43]. Calculated interface, intramolecular, and total dipole moments for the $\alpha_{(001)}$, $\dot{\alpha}_{(001)}$ and $\delta_{(001)}$ phases are listed in Table 1.

Charge transfer between the substrate and the molecular layer can be defined as the maximum absolute value of the cumulative charge transfer function $Q(z)$. [44] This function represents the total charge transferred below a plane at position *z:*

$$Q(z) = \iiint_{-\infty}^{z} \Delta\rho(x, y, z)\mathrm{d}x\mathrm{d}y\mathrm{d}z. \quad (4)$$

The optimized structures, averaged electron density differences after adsorption, and electronic charge transfer functions for the $\alpha_{(001)}$, $\dot{\alpha}_{(001)}$ and $\delta_{(001)}$ phases are depicted in Figure 4. As shown in Figure 4d, ~0.1 electrons from the $\alpha_{(001)}$ phase are transferred to the silver substrate, which induces the interface dipole moment of $\mu_{\mathrm{int}} = -1.16$ D while the molecule stays planar and we observe only a small contribution from the intramolecular dipole moment, $\mu_{\mathrm{mol}} = -0.05$ D; the total dipole moment is $-1.21$ D. After deprotonation of the $\alpha_{(001)}$ phase, i.e., removal of the hydrogen atom marked in Figure 4a and the transformation to the $\dot{\alpha}_{(001)}$ phase (Figure 4b), the molecular system undergoes qualitative changes both in the structure and electronic configuration. The deprotonated carboxylic group is pushed towards the silver substrate, which causes a bending of BDA molecules (Figure 4g, h). This is accompanied by two effects: first, ~0.2 $e^-$ are transferred from the silver substrate to the molecular layer (Figure 4e), which changes the interface dipole moment to $\mu_{\mathrm{int}} = 0.98$ D. Second, the intramolecular dipole moment is decreased to $\mu_{\mathrm{mol}} = -1.01$ D due to the out-of-plane deviation of negatively charged oxygen atoms, see Table 1. Hence, the total dipole moment in the $\dot{\alpha}_{(001)}$ phase is $-0.03$ D, i.e., it is increased by 1.17 D with respect to the $\alpha_{(001)}$ phase. After the full deprotonation of the BDA molecules and the transformation to the $\delta_{(001)}$ phase (Figure 4c), both contributions to the resulting dipole moment and charge transfer functions follow the same trend. First, the charge transfer from the silver substrate increases to ~0.5 $e^-$ (Figure 4f) which also increases the interface dipole moment to $\mu_{\mathrm{int}} = 2.98$ D. Second, the BDA molecules undergo even more prominent bending (Figure 4i), which is accompanied by an additional decrease of the intramolecular dipole moment to $\mu_{\mathrm{mol}} = -2.20$ D. As a result, the total dipole moment is increased by 0.82 D with respect to the $\dot{\alpha}_{(001)}$ phase to 0.79 D. This change is smaller by 30% than the one observed during the transformation of the $\alpha_{(001)}$ to the $\dot{\alpha}_{(001)}$ phase due to a smaller increase of the interface dipole moment by 0.14 D and a larger decrease of the intramolecular dipole moment by 0.23 D. This is consistent with the work function measurements as a function of the degree of deprotonation, showing a milder slope during the transformation of $\alpha_{(001)}$ into the $\beta_{(001)}$ phase (Figure 3b). Finally, the resulting changes in the calculated work function (Table 1) are in perfect agreement with the experiment (cf. Figure 6): they differ by 0.08 eV, 0.06 eV, and 0.05 eV for the $\alpha_{(001)}$, $\dot{\alpha}_{(001)}$ and $\delta_{(001)}$ phases, respectively.

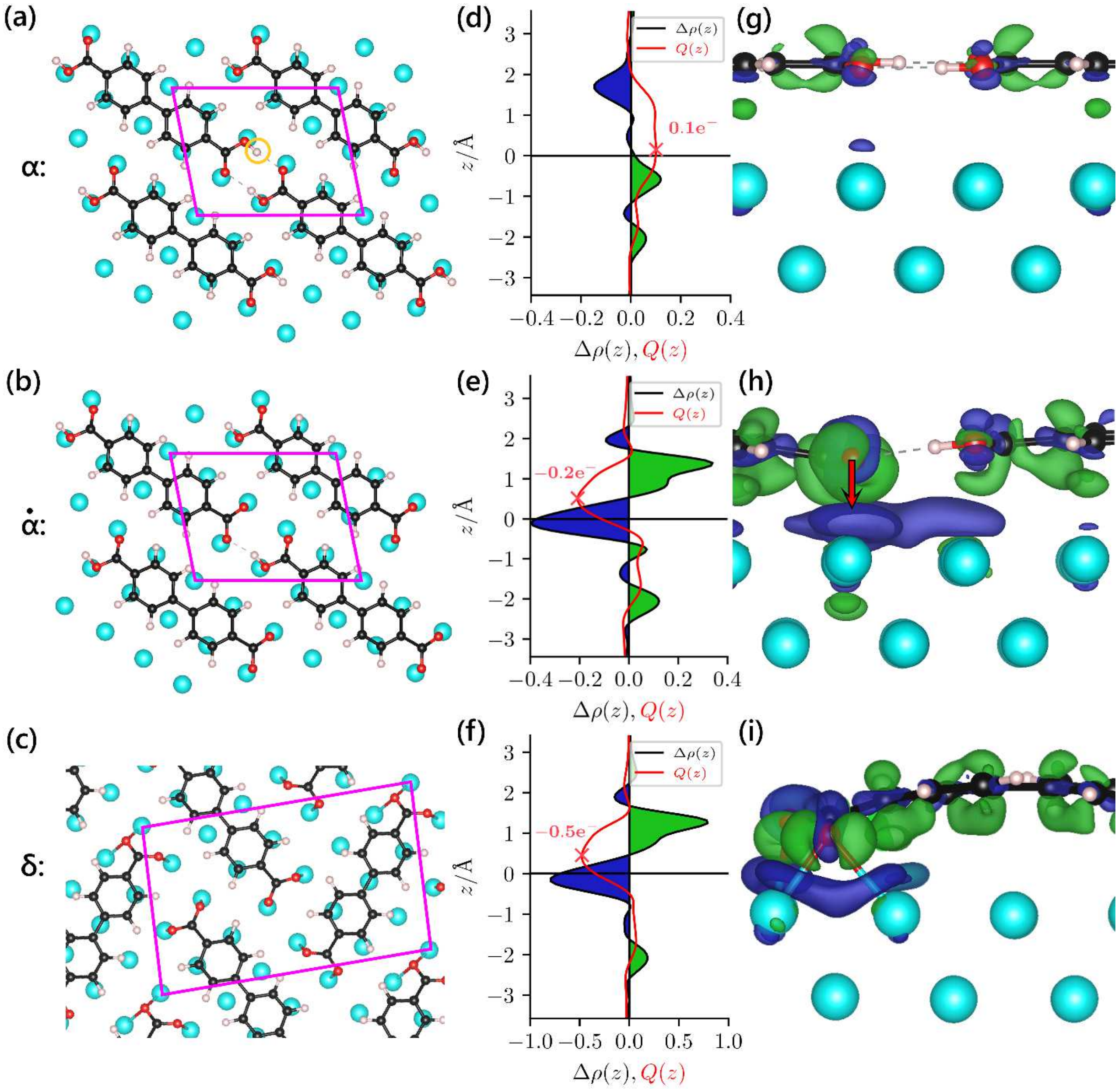

**Figure 4:** (a–c) Calculated unit cells for the α, α̇ and $\delta$ phases. Cyan, black, red, and white spheres represent Ag, C, O, and H atoms, respectively. The orange circle marks an extra hydrogen atom in the α phase. (d-f) Averaged electron density difference over the *xy* plane and charge transfer functions for the α, α̇ and δ phases, respectively. The black horizontal line represents the calculated center of a dipole (see eq. 3). (g–i) Charge density redistribution in the α, α̇, and δ phases induced

by the interaction between the silver substrate and the molecular layer. The red arrow marks the shift of the deprotonated carboxylic group towards the silver substrate

**Table 1:** Calculated dipole moments and work function changes for the $\alpha_{(001)}$, $\dot{\alpha}_{(001)}$ and $\delta_{(001)}$ phases. The calculated dipole moments for the $\delta_{(001)}$ phase are normalized to the size of the $\alpha_{(001)}$ supercell.

| phase | $\mu_{\mathrm{int}}$ [D] | $\mu_{\mathrm{mol}}$ [D] | $\mu_{\mathrm{tot}}$[D] | $e\Delta\varphi$ [eV] |
|---|---|---|---|---|
| $\alpha_{(001)}$ | $-1.16$ | $-0.05$ | $-1.20$ | $-0.53$ |
| $\dot{\alpha}_{(001)}$ | 0.98 | $-1.01$ | $-0.03$ | $-0.04$ |
| $\delta_{(001)}$ | 2.98 | $-2.20$ | 0.79 | 0.33 |

### 3.1.4. Discussion of the relationship between work function and core-level shifts in the CIL

During the thermal treatment of a BDA sub-monolayer, we have established that the work function increases linearly with the degree of deprotonation of BDA's carboxyl groups, as detailed above. The linear dependence can also be established between the position of the CIL C 1s peak and the measured WF of the full molecular layer, as given in Figure 5. On the surface of a metal, the change of the work function $e\Delta\varphi$ is induced by the formation of molecular dipoles $e\Delta\varphi_{\mathrm{Dip}}$ [43]:

$$e\Delta\varphi = e\Delta\varphi_{\mathrm{Dip}}. \qquad (5)$$

The term $e\Delta\varphi_{\mathrm{Dip}}$ comprises contributions from the interface and intramolecular dipoles. The deposited molecules induce the interfacial charge redistribution (push-back effect) that decreases the work function [10,11,45,46], whereas the deprotonated carboxyl groups display a negative charge localized on the O atoms [27,40], which results in a WF increase [47–50]. There are two contributions to the formed dipoles revealed by our DFT calculations: the interface dipoles increase the WF, whereas the intramolecular dipoles formed due to BDA molecule bending decrease the WF. Our experiments show that the $\alpha_{(001)}$ CIL displays a work function 0.61 eV lower than the Ag(001) substrate, the $\dot{\alpha}_{(001)}$ CIL's work function is similar to the substrate (0.02 eV higher), and the $\delta_{(001)}$ CIL has a WF 0.28 eV higher than the bare Ag(001) substrate, consistent with our DFT calculations. In line with the WF change, the core energy levels of BDA molecules shift accordingly as the dipoles electrostatically induce shifts in the kinetic energies of the photoelectrons [13]. Our observation is consistent with the studies on OSC bilayers in which the local electrostatic potential of formed interface dipoles was identified as the origin of the change in molecular orbital energies [22,51].

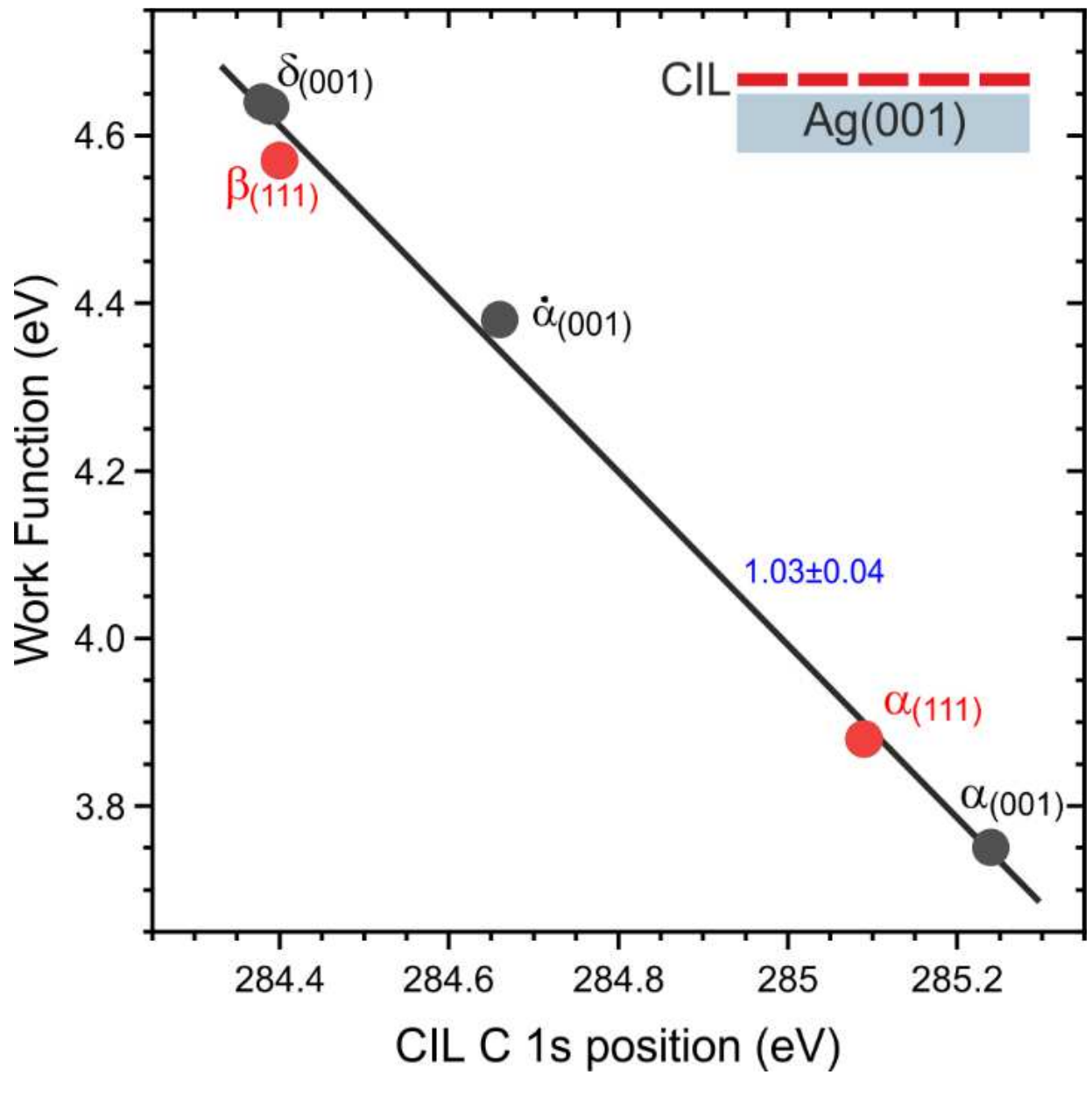

**Figure 5**: Dependence of work function on the position of the C 1s component associated with the phenyl rings for the BDA CILs marked at each point.

### 3.2 Energy level alignment in BDA bilayers

Figures 6a–d show the photoelectron spectra of the 2nd BDA layer deposited on the $\alpha_{(001)}$ and $\dot{\alpha}_{(001)}$ CILs. The spectra of the respective CILs are given in Figure 2, and the complete set of the spectra is provided in the Supplemental Material, Sections 1 and 2. After the deposition of the 2nd layer of BDA, the photoelectron spectra are more complex (Figure 6c). For their fitting, we have introduced two additional peaks with BEs of 534.2 and 532.8 eV and 1:1 ratio of intensities that represent the $\alpha$ phase in the 2nd BDA layer. For the CIL (i.e., the first layer), we have used the previously obtained peak positions, Gaussian widths, and intensity ratios and only allowed a change in the peak intensity. The peaks with the same separation and intensity ratio also appear in the spectra of the 2nd BDA layer on top of the $\dot{\alpha}_{(001)}$ (Figure 6d) but shifted to lower BEs (533.7 and 532.30 eV). The presence of the well-defined peaks associated with the protonated carboxyl groups suggests that the 2nd layer BDA molecules are intact, i.e., non-deprotonated. The intact BDA molecules in the 2nd layer were also found for all other CILs in this study. No deprotonation in the 2nd layer was observed even on the $\delta_{(001)}$ CIL, which features only deprotonated BDA molecules, even after annealing at elevated temperatures. The 2nd BDA layer started to desorb at 75 °C and disappeared at 100 °C. Up to the desorption temperature, no changes in the peaks except for a reduction in their intensity were observed.

The C 1s spectrum of the $\alpha_{(001)}$ CIL (Figure 2a) shows the phenyl-ring-associated component at 285.25 eV. After the deposition of the BDA overlayer, the main peak significantly broadens. Taking the CIL spectra as a reference, a new component related to the 2nd layer C in the phenyl rings becomes apparent in Figure 6a. It has the same width and comparable intensity as the CIL peak but is shifted by 0.35 eV to higher BE. For the $\dot{\alpha}_{(001)}$ CIL the spectrum in Figure 6b is similar, but both CIL and 2nd layer peaks are shifted to lower BEs compared to the $\alpha_{(001)}$ CIL: the CIL peak is at 284.65 eV, and the 2nd layer peak can be found at 0.35 eV higher BE with respect to the CIL. Remarkably, for both the $\alpha_{(001)}$ and $\dot{\alpha}_{(001)}$ CILs, the shift of the 2nd layer peak relative to the CIL one is the same. The BDA C 1s core level positions are summarized for all CILs in Figure 7.

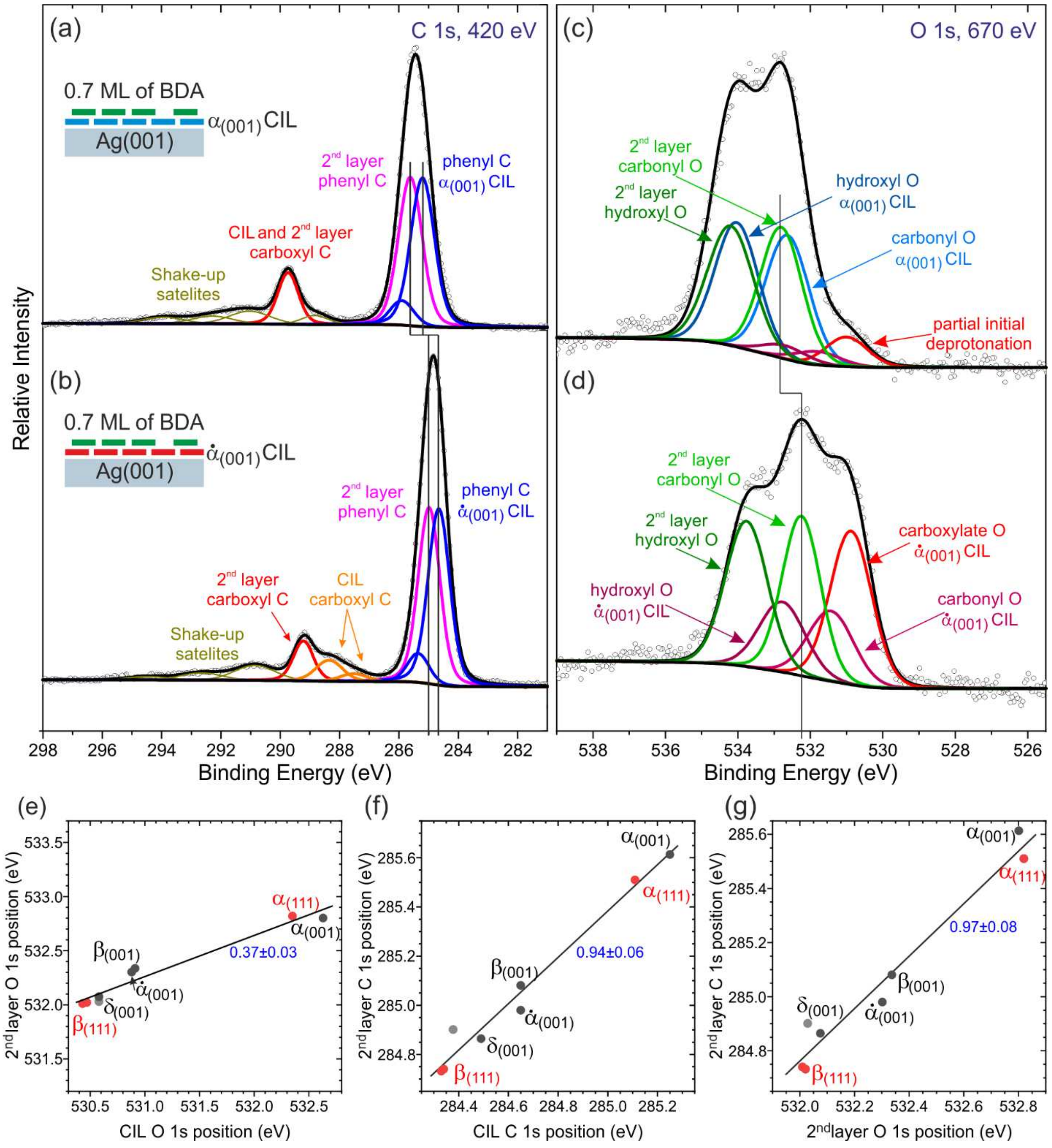

**Figure 6:** Photoelectron spectroscopy analysis of the 2nd layer of BDA on the CILs. (a, b) C 1s spectra measured by synchrotron radiation with an energy of 420 eV from samples comprising 0.7 ML of BDA deposited on the $\alpha_{(001)}$ and $\dot{\alpha}_{(001)}$ CILs. The peak components are marked accordingly. The full description and peak parameters are given in the Supplemental Material,

Section 1 and 2. (c, d) O 1s spectra (670 eV) of 0.7 ML of BDA deposited on the $\alpha_{(001)}$ and $\dot{\alpha}_{(001)}$ CILs. The vertical lines mark the shift of the peaks discussed in the text. (e, f, g) Positions of the 2nd layer peaks as a function of the CIL carbonyl O 1s position (e), the CIL phenyl C 1s position (f), and the 2nd layer O 1s position (g). The gray point for the $\delta_{(001)}$ phase marks an incompletely formed layer. The energy scale is the same on both axes in each panel. The blue value denotes the slope of the linear fit (black line).

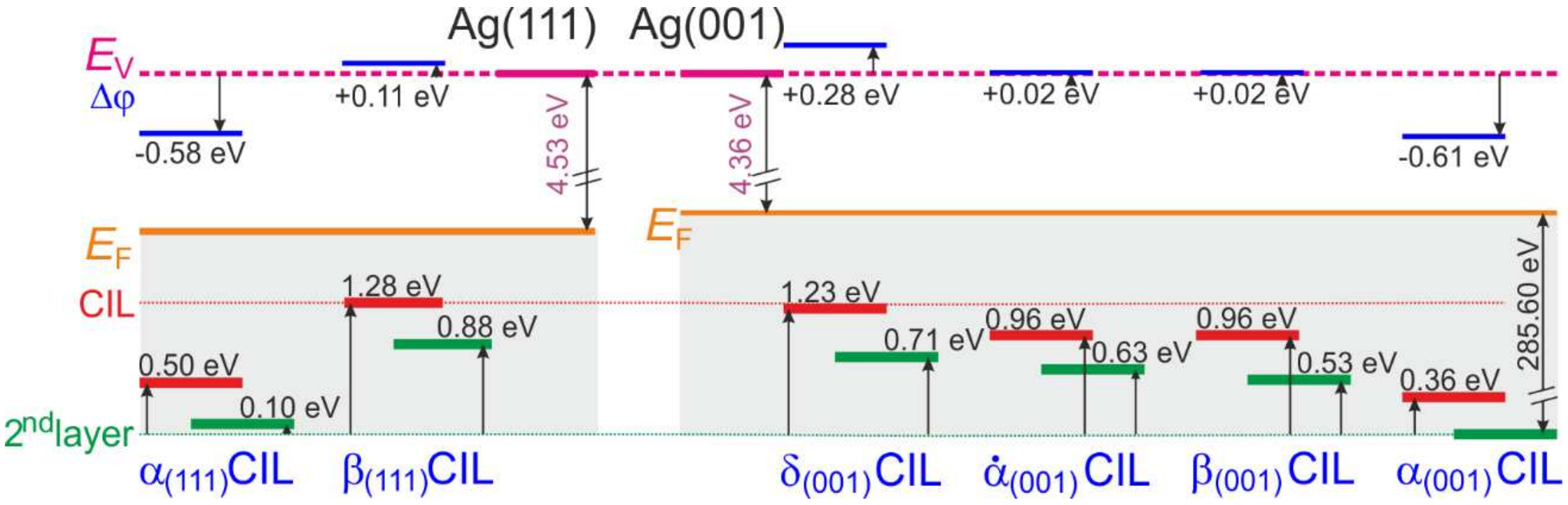

**Figure 7:** Summary of energy level positions determined for the studied CILs in an energy level diagram: vacuum level $E_{\mathrm{V}}$, Fermi level $E_{\mathrm{F}}$, C 1s core level positions of BDA in the CIL and the 2nd layer, and WF shifts $\Delta\varphi$ induced by the CILs are shown. Blue lines present the measured work functions with respect to the bare substrate (mean literature values [52] are taken for the substrate). The red and green lines mark the positions of the C 1s peak of BDA in the CILs (red lines) and 2nd layers (green lines).

Figures 6e and f compare the energy shifts of the carbonyl O 1s and C 1s peaks in the 2nd layer and the CIL. The O 1s peak components shift differently in the CIL and the 2nd layer (Figure 6e). This indicates a strong interaction of carboxylate oxygens with the substrate [46], which is accompanied by a significant electronic transfer, as shown by our DFT calculations given above. In contrast, the C 1s peak component associated with the phenyl rings shows a comparable shift in both CIL and the 2nd layer (Figure 6f; the slope is close to 1). The rigid shift hints at the absence of the chemical interaction of C species with the substrate. Finally, Figure 6g shows that the O 1s and C 1s peaks in the 2nd BDA layer exhibit the same BE shift; this implies no chemical interaction of the 2nd layer molecules with all CILs. The position of the energy levels of the 2nd layer BDA in the valence band region follows a similar trend: their shift is consistent with the shift of the core levels, as shown in Supplemental Material, Section 4. The same shift of core and valence band levels occurs when the molecular levels are aligned to the vacuum level [46,53]. In this case, we can employ the core level shifts to probe local electric dipoles that are responsible for the ELA of frontier orbitals [13]. We note that with an increasing overlayer thickness, the overlayer peaks shift towards the bulk values as given in Supplemental Material, Section 5; this behavior is probably a consequence of a decreasing core-hole screening for ionized atoms at larger distances from the metal substrate [10,20,42]. The observed shifts in the 2nd layer energy levels in the range of 0.8 eV indicate that the considered BDA molecular phases can act as CILs.

### 3.3 Continuous tuning of energy level alignment

Finally, we present a continuous tuning of the 2nd layer energy level position, which is enabled by increasing the dipole density $n_{\mathrm{Dip}}$ during the continuous isostructural phase transformation of the $\alpha_{(001)}$ to the $\dot{\alpha}_{(001)}$ CIL [27]. This transition can be achieved by sample annealing at 70 °C [27] or by secondary electrons during exposure to X-ray radiation. [40] As the 2nd layer BDA already desorbs at 70 °C, the second option is favorable and enables a simultaneous XPS analysis. The X-ray irradiation-induced deprotonation is restricted to the molecular layer that is in immediate contact with the Ag surface. Hence, in the case of 2 ML of BDA, only BDA molecules in the first layer will slowly deprotonate ($\mathcal{D}\sim 0.5$ was reached in 42 hours of irradiation) and the $n_{\mathrm{Dip}}$ will continuously increase. Snapshots from the series of XPS spectra measured during irradiation are presented in Figure 8. To evaluate the spectra, we have used the peak parameters obtained from the pure phases described above and assumed that the 2nd layer BDA does not chemically change during the irradiation. This is substantiated by (i) the absence of the deprotonation in the 2nd layer on the $\delta_{(001)}$ CIL even after annealing up to the desorption of the 2nd layer and (ii) the fact that the bonding of the carboxylate to the substrate makes the deprotonation energetically favorable. Therefore, we have kept the intensity and Gaussian widths of the 2nd layer peaks constant during the fitting of the whole series. The positions, intensity ratios, and Gaussian widths of individual components were kept constant for both CIL phases. The only free fitting parameters were the joint intensities of the α phase CIL and the $\dot{\alpha}$ phase CIL peaks and position of the 2nd layer peaks. The degree of deprotonation was determined as the ratio of the carboxylate peak intensity to the total intensity of the 1st layer (CIL) peaks.

The graph of the O 1s position as a function of $\mathcal{D}$ is given in Figure 9a. The 2nd layer energy level position upshifts linearly by 0.4 eV for $n_{\mathrm{Dip}}$ up to 0.6 nm$^{-2}$ ($\mathcal{D} = 0.25$), beyond which it slows down: an additional shift by 0.1 eV is obtained for $n_{\mathrm{Dip}} = 1.25$ nm$^{-2}$ ($\mathcal{D} = 0.5$). Figure 9b shows the work function change determined from the shift of the 2nd layer O 1s peak as a function of the density of the interface dipoles that change from 0 to 1.25 nm$^{-2}$. The shift in energy levels is given relatively to the initial 2nd layer O 1s peak position of 532.99 eV. We note that during synchrotron radiation measurements, such damage was not observed because the sample was irradiated for much shorter times (~30 minutes) with a comparable photon flux. To confirm minimal beam damage, we occasionally shifted the sample by a few millimeters and compared the spectra from the fresh and long-illuminated spots; they looked identical in all cases.

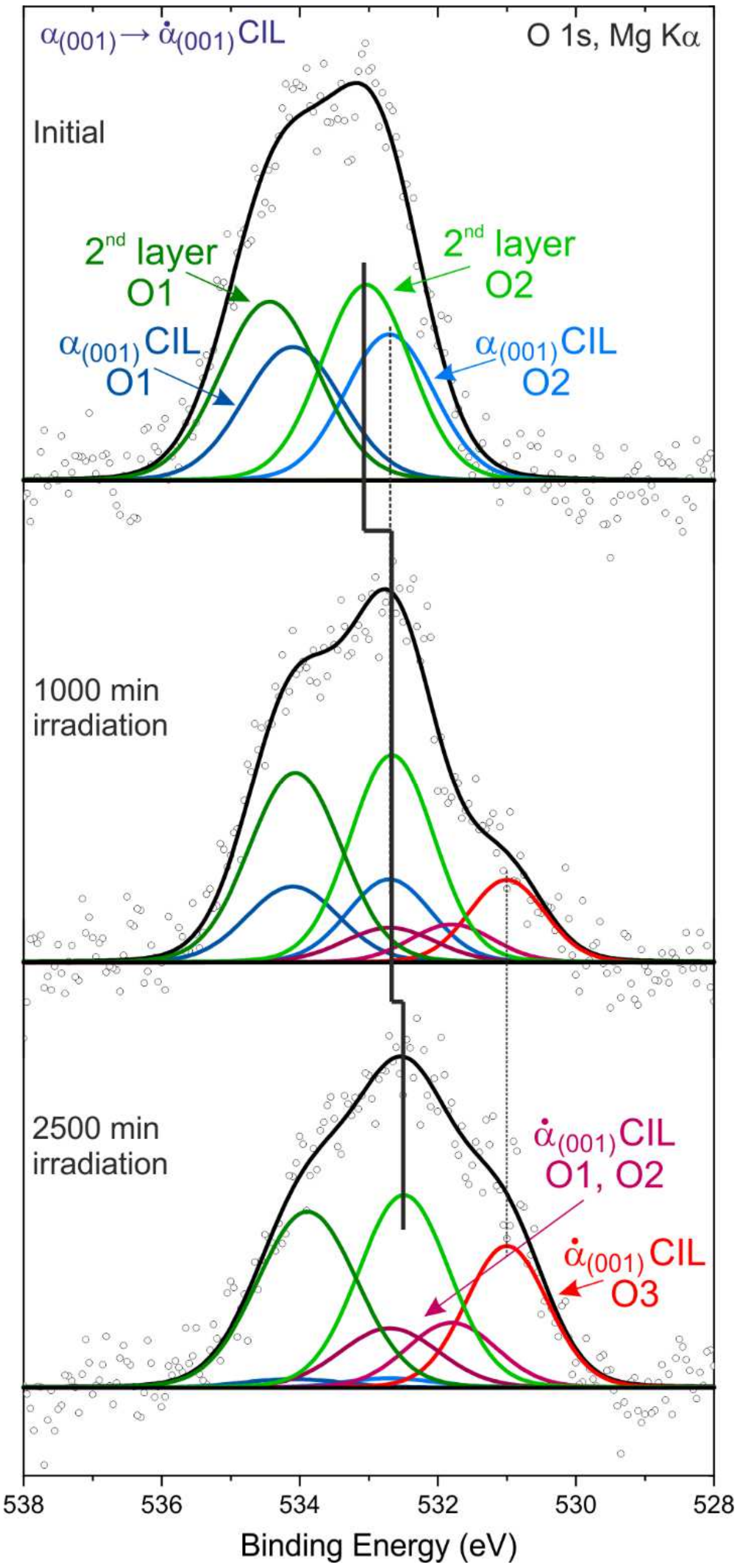

**Figure 8**: Fitting of the O 1s XPS spectra measured during 2500-min long X-ray exposure. The peaks associated with the $\alpha_{(001)}$ and $\dot{\alpha}_{(001)}$ CIL (the 1st layer) and the 2nd layer are marked accordingly.

As discussed earlier, the upshift of electronic levels in the 2nd layer is caused by the dipole component of the work function change $e\Delta\varphi_{\mathrm{Dip}}$. In line with previous works on weakly-interacting molecules that carry an electric dipole [48–50], the dependence of $e\Delta\varphi_{\mathrm{Dip}}$ on $n_{\mathrm{Dip}}$ can be approximated by the Topping model [43] Assuming that the molecular dipoles are uniformly arranged and localized at a single point, the change in the work function can be expressed as:

$$e\Delta\varphi_{\mathrm{Dip}} = \frac{e}{\varepsilon_0}\mu n_{\mathrm{Dip}}\left(1 + \frac{9\alpha}{4\pi\varepsilon_0}n_{\mathrm{Dip}}^{3/2}\right)^{-1}, \qquad (6)$$

where $\mu$ is the dipole moment, $\alpha$ is the polarizability of a surface-molecule complex, and $\varepsilon_0$ is the permittivity of the vacuum. The modeling of the measured data is given in Figure 9b. The fit of the low dipole density part, in which it is sufficient to take only the linear part of (6), i.e., $\frac{e}{\varepsilon_0}\mu n_{\mathrm{Dip}}$, gives the value of the electric dipole moment $\mu$ = (1.4 ± 0.2) D. Employing the value for the full model gives the polarizability volume $\frac{\alpha}{4\pi\varepsilon_0}$ = (3 ± 1) · $10^{-29}$ $m^3$; both values are consistent with those of similar systems [48,49]. Our DFT calculations give a value of dipole moment change between the $\alpha_{(001)}$ and the $\dot{\alpha}_{(001)}$ CILs of 1.2 D, in good agreement with the experimental value. Applying the model and derived values for the $\delta_{(001)}$ phase gives an energy shift of (0.64 ± 0.08) eV consistent with the measured value of 0.71 eV. The applicability of the model in eq. (6) for the energy level shifts in the 2nd layer of deposited molecules shows that the change in the energy level position in the BDA bilayer is caused by the dipolar contribution of the work function change.

The properties presented above suggest that the layers of carboxylic acids can be used as tunable CILs. However, several issues have to be targeted before their application in devices. First, their favorable properties should be demonstrated on OSC layers. Second, the long-term stability of a deprotonation state in the carboxylic layers and the interaction of the layers with ambient conditions should be evaluated. With respect to the stability of a deprotonation state, our recent work shows that it is stabilized in specific molecular phases [29]. The other possibility is to design molecules possessing required properties in the fully deprotonated state.

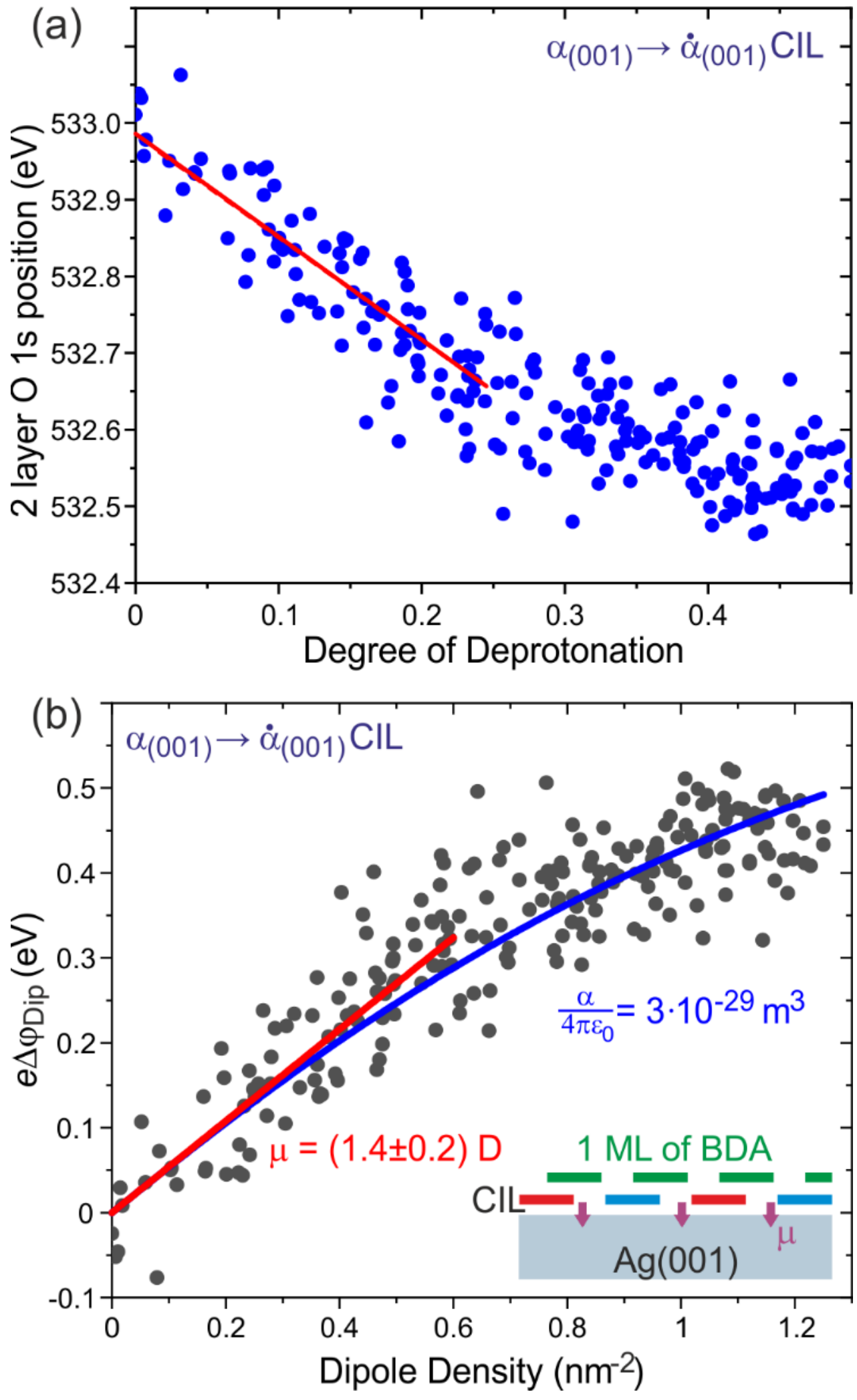

**Figure 9**: (a) The shift of the 2nd layer O 1s peak as a function of the degree of deprotonation of the CIL. The variable degree of deprotonation was achieved during the isostructural $\alpha_{(001)}$ to $\dot{\alpha}_{(001)}$ transformation. (b) The work function change determined from the shift of the 2nd layer O 1s peak as a function of electric dipole density $n_{\mathrm{Dip}}$ during the isostructural $\alpha_{(001)}$ to $\dot{\alpha}_{(001)}$

transformation. The red line is a linear fit of the initial region, and the blue curve is the full model given by Eq. (6).

## 4. Conclusions

In conclusion, we have employed a single molecule to prepare a range of CILs on Ag(001) and Ag(111) surfaces by controlled deprotonation of the carboxylic groups, which results in the formation of interface dipoles with densities up to 2.3 $nm^{-2}$. With an increasing density of dipoles, the sample WF increases by up to 0.8 eV. We have identified two distinct contributions to the formed dipoles: the interface dipoles increase the WF, whereas the intramolecular dipoles formed due to BDA molecule bending decrease the WF. The core energy levels of BDA molecules shift linearly with the WF change as the dipoles electrostatically induce shifts in the kinetic energies of the photoelectrons. We have demonstrated a continuous shift of energy levels of molecular overlayers in the range of 0 to 0.4 eV. This possibility makes the presented system a prospective candidate for the realization of efficient CILs in electronic devices featuring OSC. The presented CILs feature flat-lying phenyl rings electronically strongly coupled to the metal substrate, which could serve as a favorable template for the growth of OSCs in the bulk-like structure.

ASSOCIATED CONTENT

Supplemental Material: (1) Analysis of the photoelectron spectra of the BDA CILs on Ag(001); (2) Analysis of the photoelectron spectra of the BDA CILs on Ag(111); (3) Work function measurement of the CILs on Ag(001) and Ag(111) substrates; (4) Mutual position of the valence and O 1s peaks in the $2^{nd}$ layer; (5) Position of the overlayer peaks as a function of the overlayer thickness.

AUTHOR'S INFORMATION

**Author's Contributions**

P.P. provided the initial idea. P.P. and J.Č. conceptualized the research. V.S. prepared the samples and performed all laboratory investigations. V.S., P.P. A.M., T.S., M.B., and J.Č. performed the synchrotron radiation experiments. J.P. and A.S. did DFT calculations. J.Č. developed the methodology, supervised the experiments, analyzed and interpreted the data, and wrote the initial manuscript. All authors discussed the results and contributed to the final manuscript.

**Competing interests**

The authors declare no competing financial interests.

ACKNOWLEDGMENTS

This research has been supported by GAČR, project No. 22-04551S. CzechNanoLab project LM2018110 funded by MEYS CR is gratefully acknowledged for the financial support of the measurements at CEITEC Nano Research Infrastructure. The authors acknowledge the CERIC-ERIC Consortium for access to experimental facilities and financial support.

DATA AVAILABILITY

The data that support the findings of this study are available from the corresponding author upon reasonable request.

REFERENCES

[1] S. Zou, Y. Shen, F. Xie, J. Chen, Y. Li, and J.-X. Tang, *Recent Advances in Organic Light-Emitting Diodes: Toward Smart Lighting and Displays*, Mater. Chem. Front. **4**, 788 (2020).

[2] Y. Huang, E.-L. Hsiang, M.-Y. Deng, and S.-T. Wu, *Mini-LED, Micro-LED and OLED Displays: Present Status and Future Perspectives*, Light Sci. Appl. **9**, 105 (2020).

[3] S. Wang, H. Zhang, B. Zhang, Z. Xie, and W. Wong, *Towards High-Power-Efficiency Solution-Processed OLEDs: Material and Device Perspectives*, Mater. Sci. Eng. R Reports **140**, 100547 (2020).

[4] S. K. Park, J. H. Kim, and S. Y. Park, *Organic 2D Optoelectronic Crystals: Charge Transport, Emerging Functions, and Their Design Perspective*, Adv. Mater. **30**, 1704759 (2018).

[5] J. Gao, J. Wang, C. Xu, Z. Hu, X. Ma, X. Zhang, L. Niu, J. Zhang, and F. Zhang, *A Critical Review on Efficient Thick-Film Organic Solar Cells*, Sol. RRL **4**, 2000364 (2020).

[6] Y. Yu, Q. Ma, H. Ling, W. Li, R. Ju, L. Bian, N. Shi, Y. Qian, M. Yi, L. Xie, and W. Huang, *Small-Molecule-Based Organic Field-Effect Transistor for Nonvolatile Memory and Artificial Synapse*, Adv. Funct. Mater. **29**, 1904602 (2019).

[7] M. Waldrip, O. D. Jurchescu, D. J. Gundlach, and E. G. Bittle, *Contact Resistance in Organic Field-Effect Transistors: Conquering the Barrier*, Adv. Funct. Mater. **30**, 1 (2020).

[8] H. Klauk, *Will We See Gigahertz Organic Transistors?*, Adv. Electron. Mater. **4**, 1700474 (2018).

[9] N. Koch, *Opportunities for Energy Level Tuning at Inorganic/Organic Semiconductor Interfaces*, Appl. Phys. Lett. **119**, 260501 (2021).

[10] A. Franco-Cañellas, S. Duhm, A. Gerlach, and F. Schreiber, *Binding and Electronic Level Alignment of π -Conjugated Systems on Metals*, Reports Prog. Phys. **83**, 066501 (2020).

[11] R. Otero, A. L. Vázquez de Parga, and J. M. Gallego, *Electronic, Structural and Chemical Effects of Charge-Transfer at Organic/Inorganic Interfaces*, Surf. Sci. Rep. **72**, 105 (2017).

[12] M. Fahlman, S. Fabiano, V. Gueskine, D. Simon, M. Berggren, and X. Crispin, *Interfaces in Organic Electronics*, Nat. Rev. Mater. **4**, 627 (2019).

[13] E. Zojer, T. C. Taucher, and O. T. Hofmann, *The Impact of Dipolar Layers on the Electronic Properties of Organic/Inorganic Hybrid Interfaces*, Adv. Mater. Interfaces **6**, 1900581 (2019).

[14] S. Braun, W. R. Salaneck, and M. Fahlman, *Energy-Level Alignment at Organic/Metal and Organic/Organic Interfaces*, Adv. Mater. **21**, 1450 (2009).

[15] O. T. Hofmann, P. Rinke, M. Scheffler, and G. Heimel, *Integer versus Fractional Charge Transfer at Metal(/Insulator)/Organic Interfaces: Cu(/NaCl)/TCNE*, ACS Nano **9**, 5391 (2015).

[16] M. P. Das, F. Green, S. M. Bose, S. N. Behera, and B. K. Roul, *Dissipation in a Quantum Wire: Fact and Fantasy*, in *AIP Conference Proceedings*, Vol. 1063 (AIP, 2008), pp. 26–34.

[17] H. Chen, W. Zhang, M. Li, G. He, and X. Guo, *Interface Engineering in Organic Field-*

*Effect Transistors: Principles, Applications, and Perspectives*, Chem. Rev. **120**, 2879 (2020).

[18] K.-G. Lim, S. Ahn, and T.-W. Lee, *Energy Level Alignment of Dipolar Interface Layer in Organic and Hybrid Perovskite Solar Cells*, J. Mater. Chem. C **6**, 2915 (2018).

[19] E. Goiri, P. Borghetti, A. El-Sayed, J. E. Ortega, and D. G. de Oteyza, *Multi-Component Organic Layers on Metal Substrates*, Adv. Mater. **28**, 1340 (2016).

[20] J. L. Zhang, X. Ye, C. Gu, C. Han, S. Sun, L. Wang, and W. Chen, *Non-Covalent Interaction Controlled 2D Organic Semiconductor Films: Molecular Self-Assembly, Electronic and Optical Properties, and Electronic Devices*, Surf. Sci. Rep. **75**, 100481 (2020).

[21] N. Koch, S. Duhm, J. P. Rabe, A. Vollmer, and R. L. Johnson, *Optimized Hole Injection with Strong Electron Acceptors at Organic-Metal Interfaces*, Phys. Rev. Lett. **95**, 237601 (2005).

[22] P. Borghetti, D. G. de Oteyza, C. Rogero, E. Goiri, A. Verdini, A. Cossaro, L. Floreano, and J. E. Ortega, *Molecular-Level Realignment in Donor–Acceptor Bilayer Blends on Metals*, J. Phys. Chem. C **120**, 5997 (2016).

[23] R. Nouchi, M. Shigeno, N. Yamada, T. Nishino, K. Tanigaki, and M. Yamaguchi, *Reversible Switching of Charge Injection Barriers at Metal/Organic-Semiconductor Contacts Modified with Structurally Disordered Molecular Monolayers*, Appl. Phys. Lett. **104**, 013308 (2014).

[24] P. Hurdax, M. Hollerer, P. Puschnig, D. Lüftner, L. Egger, M. G. Ramsey, and M. Sterrer,

*Controlling the Charge Transfer across Thin Dielectric Interlayers*, Adv. Mater. Interfaces **7**, 2000592 (2020).

[25] H. Edlbauer, E. Zojer, and O. T. Hofmann, *Postadsorption Work Function Tuning via Hydrogen Pressure Control*, J. Phys. Chem. C **119**, 27162 (2015).

[26] A. Kahn, *Fermi Level, Work Function and Vacuum Level*, Mater. Horizons **3**, 7 (2016).

[27] P. Procházka, L. Kormoš, A. Shahsavar, V. Stará, A. O. Makoveev, T. Skála, M. Blatnik, and J. Čechal, *Phase Transformations in a Complete Monolayer of 4,4'-Biphenyl-Dicarboxylic Acid on Ag(0 0 1)*, Appl. Surf. Sci. **547**, 149115 (2021).

[28] P. Procházka, M. A. Gosalvez, L. Kormoš, B. de la Torre, A. Gallardo, J. Alberdi-Rodriguez, T. Chutora, A. O. Makoveev, A. Shahsavar, A. Arnau, P. Jelínek, and J. Čechal, *Multiscale Analysis of Phase Transformations in Self-Assembled Layers of 4,4'-Biphenyl Dicarboxylic Acid on the Ag(001) Surface*, ACS Nano **14**, 7269 (2020).

[29] A. O. Makoveev, P. Procházka, M. Blatnik, L. Kormoš, T. Skála, and J. Čechal, *Role of Phase Stabilization and Surface Orientation in 4,4'-Biphenyl-Dicarboxylic Acid Self-Assembly and Transformation on Silver Substrates*, J. Phys. Chem. C **126**, 9989 (2022).

[30] L. Kormoš, P. Procházka, A. O. Makoveev, and J. Čechal, *Complex K-Uniform Tilings by a Simple Bitopic Precursor Self-Assembled on Ag(001) Surface*, Nat. Commun. **11**, 1856 (2020).

[31] G. Kresse and J. Furthmüller, *Efficient Iterative Schemes for Ab Initio Total-Energy Calculations Using a Plane-Wave Basis Set*, Phys. Rev. B **54**, 11169 (1996).

[32] G. Kresse and D. Joubert, *From Ultrasoft Pseudopotentials to the Projector Augmented-Wave Method*, Phys. Rev. B **59**, 1758 (1999).

[33] H. J. Monkhorst and J. D. Pack, *Special Points for Brillouin-Zone Integrations*, Phys. Rev. B **13**, 5188 (1976).

[34] J. Klimeš, D. R. Bowler, and A. Michaelides, *Chemical Accuracy for the van Der Waals Density Functional*, J. Phys. Condens. Matter **22**, 022201 (2010).

[35] S. Grimme, J. Antony, S. Ehrlich, and H. Krieg, *A Consistent and Accurate Ab Initio Parametrization of Density Functional Dispersion Correction (DFT-D) for the 94 Elements H-Pu*, J. Chem. Phys. **132**, 154104 (2010).

[36] J. Klimeš, D. R. Bowler, and A. Michaelides, *Van Der Waals Density Functionals Applied to Solids*, Phys. Rev. B **83**, 195131 (2011).

[37] J. MacLeod, *Design and Construction of On-Surface Molecular Nanoarchitectures: Lessons and Trends from Trimesic Acid and Other Small Carboxlyated Building Blocks*, J. Phys. D **53**, 043002 (2020).

[38] A. Schöll, Y. Zou, M. Jung, T. Schmidt, R. Fink, and E. Umbach, *Line Shapes and Satellites in High-Resolution x-Ray Photoelectron Spectra of Large π-Conjugated Organic Molecules*, J. Chem. Phys. **121**, 10260 (2004).

[39] M. Häming, A. Schöll, E. Umbach, and F. Reinert, *Adsorbate-Substrate Charge Transfer and Electron-Hole Correlation at Adsorbate/Metal Interfaces*, Phys. Rev. B **85**, 235132 (2012).

[40] A. Makoveev, P. Procházka, A. Shahsavar, L. Kormoš, T. Krajňák, V. Stará, and J. Čechal, *Kinetic Control of Self-Assembly Using a Low-Energy Electron Beam*, Appl. Surf. Sci. **600**, 154106 (2022).

[41] T. Schultz, P. Amsalem, N. B. Kotadiya, T. Lenz, P. W. M. Blom, and N. Koch, *Importance of Substrate Work Function Homogeneity for Reliable Ionization Energy Determination by Photoelectron Spectroscopy*, Phys. Status Solidi **256**, 1800299 (2019).

[42] H. Ishii, K. Sugiyama, E. Ito, and K. Seki, *Energy Level Alignment and Interfacial Electronic Structures at Organic/Metal and Organic/Organic Interfaces*, Adv. Mater. **11**, 605 (1999).

[43] H. Lüth, *Solid Surfaces, Interfaces and Thin Films*, 6th ed. (Springer Berlin Heidelberg, 2015).

[44] O. T. Hofmann, H. Glowatzki, C. Bürker, G. M. Rangger, B. Bröker, J. Niederhausen, T. Hosokai, I. Salzmann, R.-P. Blum, R. Rieger, A. Vollmer, P. Rajput, A. Gerlach, K. Müllen, F. Schreiber, E. Zojer, N. Koch, and S. Duhm, *Orientation-Dependent Work-Function Modification Using Substituted Pyrene-Based Acceptors*, J. Phys. Chem. C **121**, 24657 (2017).

[45] A. Franco-Cañellas, Q. Wang, K. Broch, D. A. Duncan, P. K. Thakur, L. Liu, S. Kera, A. Gerlach, S. Duhm, and F. Schreiber, *Metal-Organic Interface Functionalization via Acceptor End Groups: PTCDI on Coinage Metals*, Phys. Rev. Mater. **1**, 013001(R) (2017).

[46] Q. Wang, A. Franco-Cañellas, P. Ji, C. Bürker, R.-B. Wang, K. Broch, P. K. Thakur, T.-L. Lee, H. Zhang, A. Gerlach, L. Chi, S. Duhm, and F. Schreiber, *Bilayer Formation vs*

*Molecular Exchange in Organic Heterostructures: Strong Impact of Subtle Changes in Molecular Structure*, J. Phys. Chem. C **122**, 9480 (2018).

[47] S. Kera, Y. Yabuuchi, H. Yamane, H. Setoyama, K. K. Okudaira, A. Kahn, and N. Ueno, *Impact of an Interface Dipole Layer on Molecular Level Alignment at an Organic-Conductor Interface Studied by Ultraviolet Photoemission Spectroscopy*, Phys. Rev. B **70**, 085304 (2004).

[48] H. Fukagawa, H. Yamane, S. Kera, K. K. Okudaira, and N. Ueno, *Experimental Estimation of the Electric Dipole Moment and Polarizability of Titanyl Phthalocyanine Using Ultraviolet Photoelectron Spectroscopy*, Phys. Rev. B **73**, 041302(R) (2006).

[49] H. Fukagawa, S. Hosoumi, H. Yamane, S. Kera, and N. Ueno, *Dielectric Properties of Polar-Phthalocyanine Monolayer Systems with Repulsive Dipole Interaction*, Phys. Rev. B **83**, 085304 (2011).

[50] A. Gerlach, T. Hosokai, S. Duhm, S. Kera, O. T. Hofmann, E. Zojer, J. Zegenhagen, and F. Schreiber, *Orientational Ordering of Nonplanar Phthalocyanines on Cu(111): Strength and Orientation of the Electric Dipole Moment*, Phys. Rev. Lett. **106**, 156102 (2011).

[51] M. Häming, M. Greif, C. Sauer, A. Schöll, and F. Reinert, *Electronic Structure of Ultrathin Heteromolecular Organic-Metal Interfaces: SnPc/PTCDA/Ag(111) and SnPc/Ag(111)*, Phys. Rev. B **82**, 235432 (2010).

[52] G. N. Derry, M. E. Kern, and E. H. Worth, *Recommended Values of Clean Metal Surface Work Functions*, J. Vac. Sci. Technol. A Vacuum, Surfaces, Film. **33**, 060801 (2015).

[53] A. El-Sayed, P. Borghetti, E. Goiri, C. Rogero, L. Floreano, G. Lovat, D. J. Mowbray, J. L. Cabellos, Y. Wakayama, A. Rubio, J. E. Ortega, and D. G. de Oteyza, *Understanding Energy-Level Alignment in Donor–Acceptor/Metal Interfaces from Core-Level Shifts*, ACS Nano **7**, 6914 (2013).

SUPPLEMENTAL MATERIAL

# Tunable Energy Level Alignment in the Multilayers of Carboxylic Acids on Silver

Veronika Stará,[1] Pavel Procházka,[1] Jakub Planer,[1] Azin Shahsavar,[1] Anton O. Makoveev,[1] Tomáš Skála,[2] Matthias Blatnik,[1] Jan Čechal[1,3*]

[1] CEITEC - Central European Institute of Technology, Brno University of Technology, Purkyňova 123, 612 00 Brno, Czech Republic.

[2] Department of Surface and Plasma Science, Faculty of Mathematics and Physics, Charles University, V Holešovičkách 2, 18000 Prague 8, Czech Republic

[3] Institute of Physical Engineering, Brno University of Technology, Technická 2896/2, 616 69 Brno, Czech Republic.

CONTENTS:

## 1. Analysis of the photoelectron spectra of BDA CILs on Ag(001)

The O 1s spectra were fitted by 2 – 3 Voigt components (parameters given in Table S1) using a Shirley background, as shown in Figure S1. All fits are consistent with our previous laboratory measurements using a standard, non-monochromatic X-ray source [1,2].

The spectra of the $\alpha_{(001)}$ CIL reveal slight initial deprotonation (the degree of deprotonation is 0.1). For the $\delta_{(001)}$ CIL, the spectra show a residual degree of deprotonation of 0.80 or 0.95. This is a balance between keeping the high BDA coverage while reaching a high degree of deprotonation. There is a very narrow interval of temperatures for the phase transformation, where the coverage is maintained, and significant BDA deprotonation occurs. Partial desorption is probably the reason for a lower coverage of 0.95; the particular point is highlighted in a lighter shade of gray in Figure 2 of the main text.

Concerning the C 1s spectra (Figure S2), the synchrotron radiation provides a much higher resolution than achievable in our laboratory. This enabled us to distinguish components of the phenyl-ring-associated components (C1 component), split of carboxyl related C 1s component (C2 component) in the $\dot{\alpha}_{(001)}$ and $\beta_{(001)}$ CILs, and better visualize the presence of shake-up satellites. Nevertheless, the peak-fitting parameters (Table S2) are similar to those obtained by a non-monochromatic X-ray source. The bilayer sample was fitted with parameters of the previously measured CILs fixed except for their intensity.

The spectra for the valence band region (0 – 14 eV) are shown in Figure S3. The analysis is complicated by a significant overlap of BDA-related peaks with the Ag ones. Hence, we have taken the deepest one (9.3 – 10.3 eV) for further investigation.

**Table S1**: Peak-fit parameters: binding energy position (BE) and FWHM of the Gaussian part of the O 1s peak components. Voigt function was used for fitting; the width of the Lorentzian component was 0.1 eV. The BE shifts (from the peak, marked +0.0) are given in brackets. The degree of deprotonation of carboxyl groups is provided for all the phases.

| | **Component O1 (hydroxyl oxygen)** | | **Component O2 (carbonyl oxygen)** | | **Component O3 (carboxylate oxygen)** | | **Degree of Deprotonation** |
|---|---|---|---|---|---|---|---|
| **CIL Ag(001)** | BE (eV) | FWHM (eV) | BE (eV) | FWHM (eV) | BE (eV) | FWHM (eV) | |
| **α** | 534.05 (+1.4) | 1.4 | 532.65 (+0.0) | 1.4 | – | – | 0.0 |
| **ἀ** | 532.80 (+1.35) | 1.4 | 531.45 (+0.0) | 1.4 | 530.90 (−0.55) | 1.3 | 0.5 |
| **β** | 532.80 (+1.35) | 1.4 | 531.45 (+0.0) | 1.4 | 530.90 (−0.55) | 1.3 | 0.5 |
| **δ** | | | | | 530.55 | 1.1 | 1.0 |

The position of the Ag $3d_{5/2}$ peak was in the interval of (368.20 ± 0.02) eV for all the measurements, which confirms that the energy tuning reproducibility is much better than the spectral resolution.

**Table S2**: Peak-fit parameters: binding energy position (BE) and FWHM of the Gaussian part of the C 1s peak components. Voigt function was used for fitting; the width of the Lorentzian component was 0.2 eV. The BE shifts (from the peak, marked +0.0) are given in brackets. The ratio of the C2 and C1 intensities match the theoretical one (2/12 = 0.167). A shoulder for the C1 component was necessary to obtain a good fit for CIL; [1] it was shifted by 0.7 eV to higher binding energy, its intensity was 0.165 of the main component, and it had the same Gaussian width. It models the weak asymmetry of the main peak, and since its parameters are constrained, it does not represent any additional degree of fitting freedom.

| | **Component C1 (phenyl rings)** | | **Component C2 (carboxyl carbon)** | | **C2/C1** |
|---|---|---|---|---|---|
| **CIL Ag(001)** | BE (eV) | FWHM (eV) | BE (eV) | FWHM (eV) | |
| **α** | 285.25 (+0.0) | 1.0 | 289.65 (+4.4) | 0.7 | 0.14 |
| **ἀ** | 284.65 (+0.0) | 0.8 | 288.35 and 287.50 (+3.70 and +2.85) | 0.9 | 0.17 |
| **β** | 284.65 (+0.0) | 0.8 | 288.50 and 287.70 (+3.85 and +3.05) | 1.0 | 0.17 |
| **δ** | 284.50 (+0.0) | 0.8 | 287.50 (+3.00) | 1.2 | 0.14 |

The position of the Ag $3d_{5/2}$ peak was in the interval of (368.20 ± 0.02) eV for all the measurements, which confirms that the energy tuning reproducibility is much better than the spectral resolution.

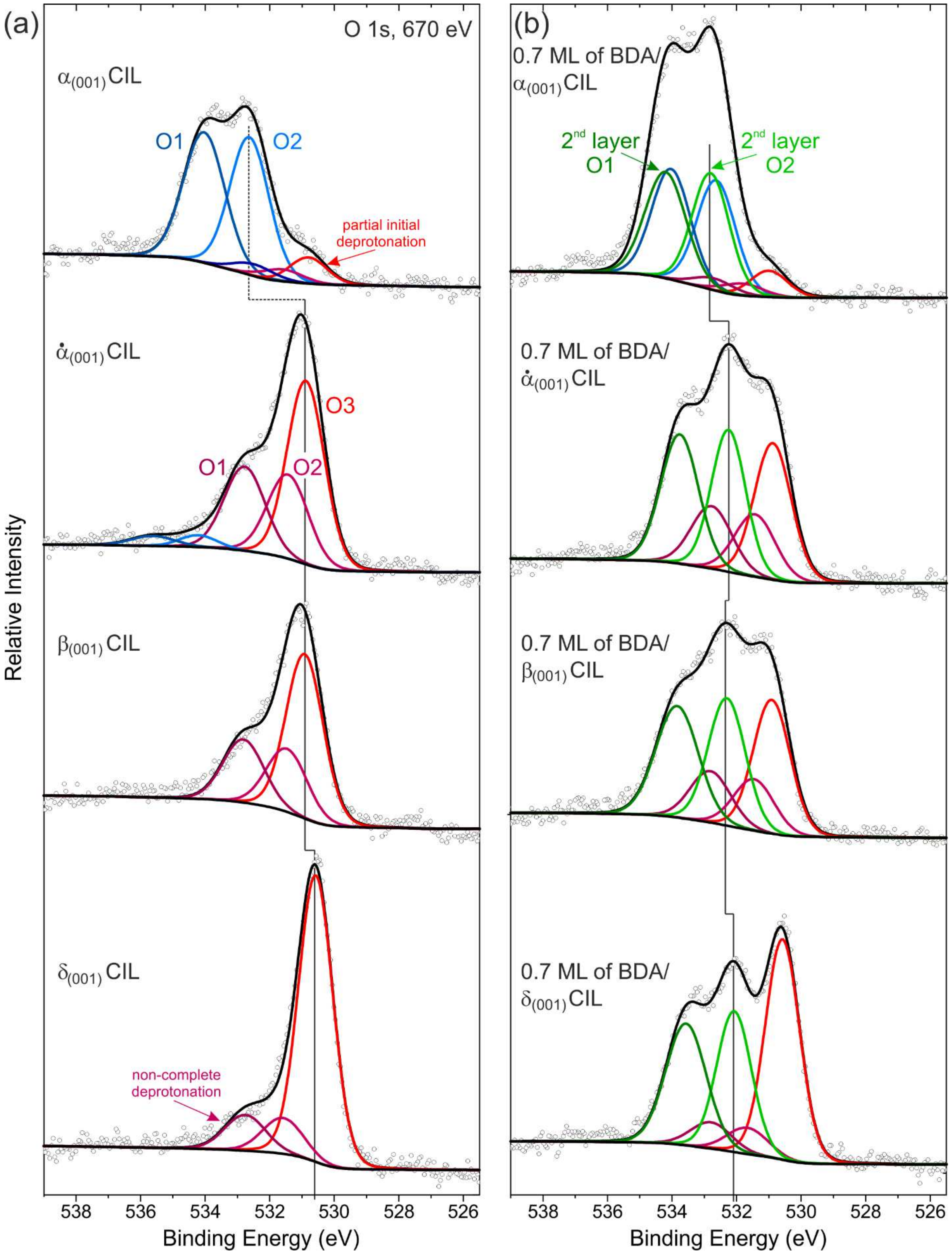

**Figure S1**: O 1s spectra of (a) CILs on Ag(001) surface and (b) 0.7 ML of BDA on the CILs on Ag(001). The vertical lines mark the peak positions used to describe CIL properties.

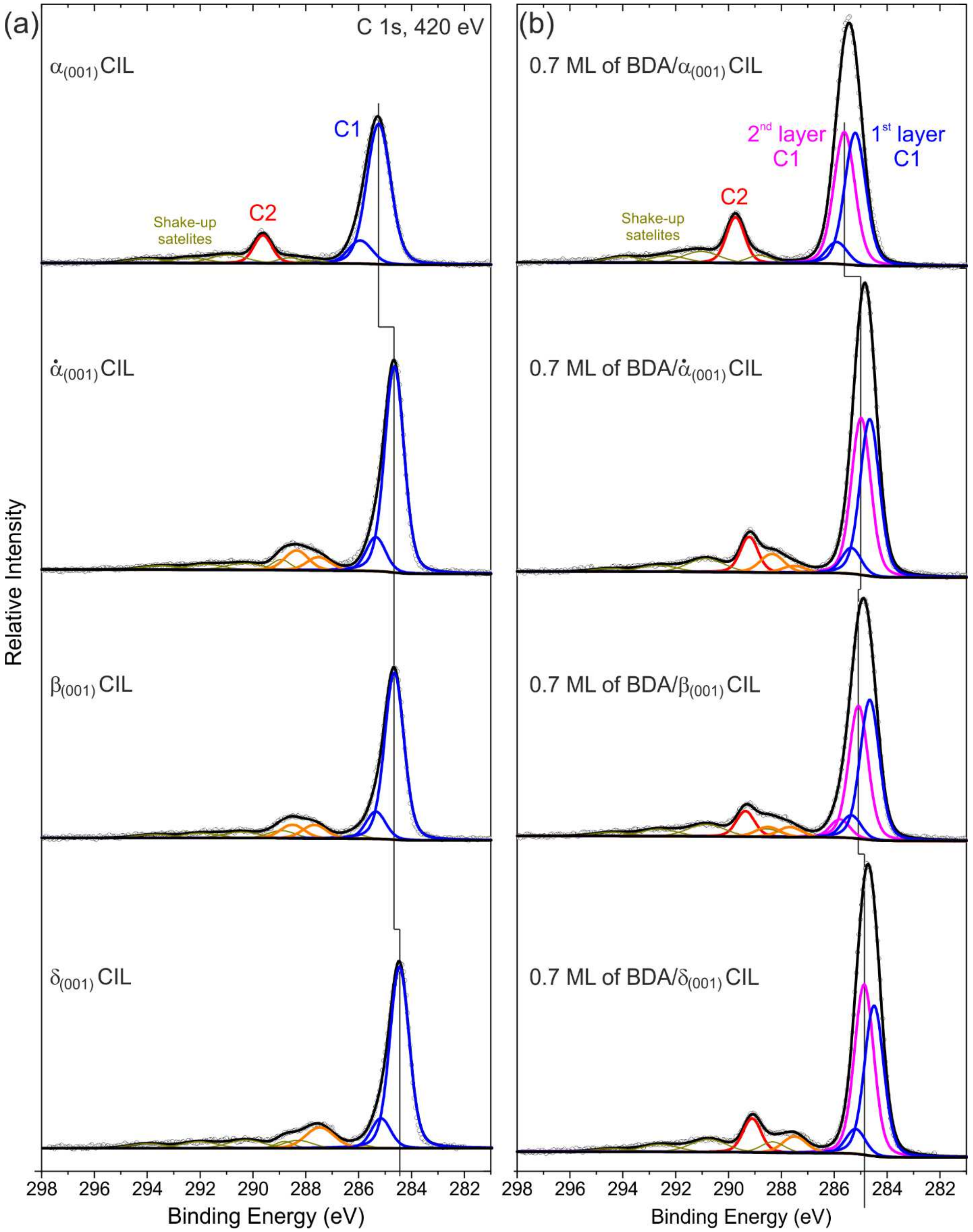

**Figure S2**: C 1s spectra of (a) CILs on Ag(001) surface and (b) 0.7 ML of BDA deposited on the CILs. The vertical lines mark the peak positions used to describe CIL properties.

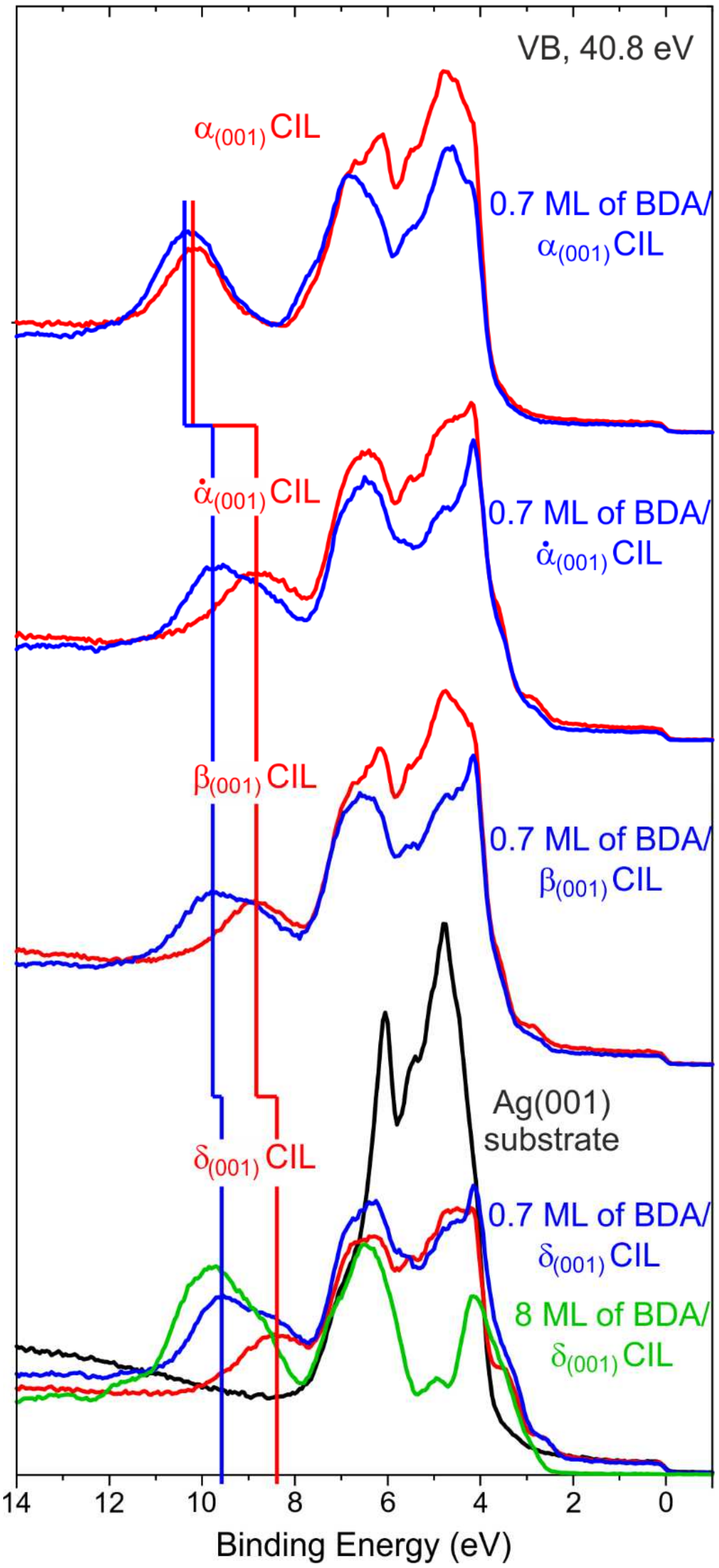

**Figure S3**: Valence band spectra of Ag(001) surface (black), CILs (red), and 0.7 ML of BDA on the CILs (blue). The spectrum of 8 ML of BDA (green) is given for reference of BDA energy levels (thick BDA layer on the $\delta_{(001)}$ CIL).

## 2. Analysis of the photoelectron spectra of BDA CILs on Ag(111)

The fitting of spectra associated with the BDA layers on both α and β phase CILs is given in Figure S4; we have employed the previously reported C 1s and O 1s peak parameters. [6] The measured valence band spectra are given in Figure S5.

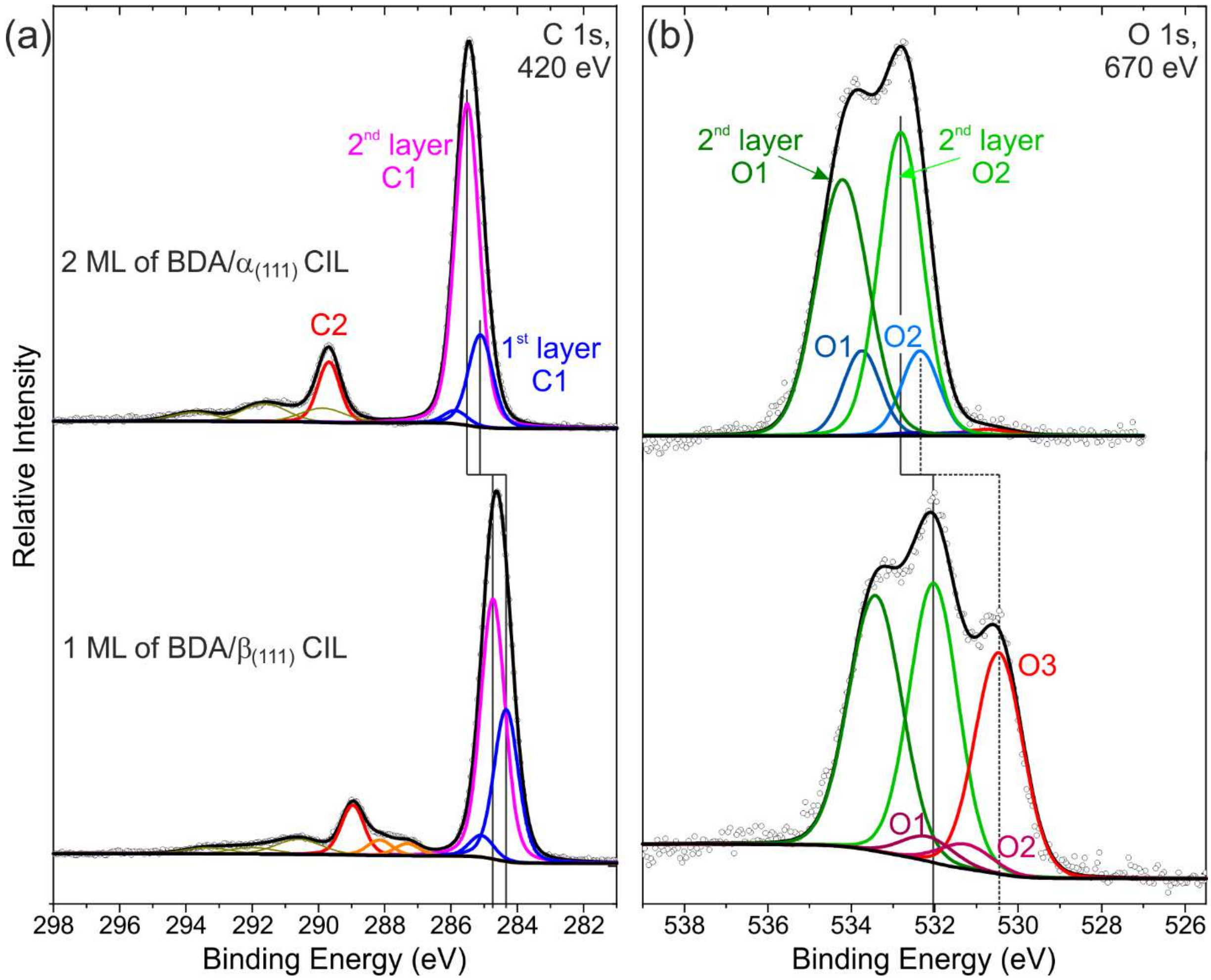

**Figure S4**: C 1s (a) and O 1s (b) spectra of BDA deposited on CILs on Ag(111) surface. The vertical lines mark the peak positions used to describe CIL properties.

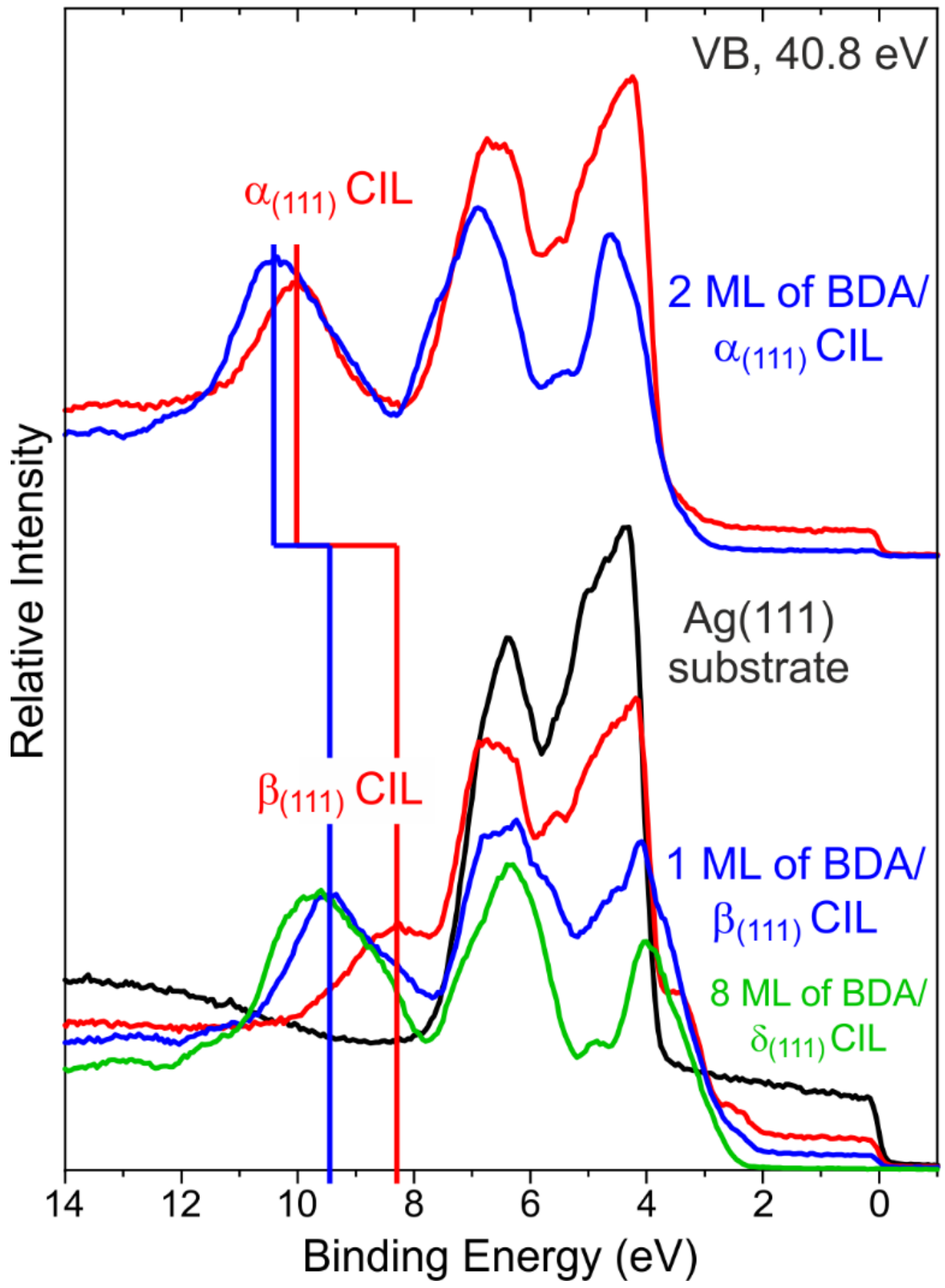

**Figure S5**: Valence band spectra of Ag(111) surface (black), CILs (red), and BDA on the CILs (blue). The spectrum of 8 ML of BDA (green) is given to provide a reference of BDA energy levels (thick BDA layer on $\beta_{(111)}$ CIL).

## 3. Work function measurement of CILs on Ag(001) and Ag(111) substrates

We have determined the work function for all considered CILs and bare substrates employing a standard procedure of measuring secondary electron cut-off (SEC) and Fermi levels as $\varphi = h\nu - E_{\mathrm{SEC}}$, where $h\nu$ is the energy of employed radiation (here, 40.77 eV) and $E_{\mathrm{SEC}}$ the measured binding energy of SEC. The measured spectra are given in Figure S6.

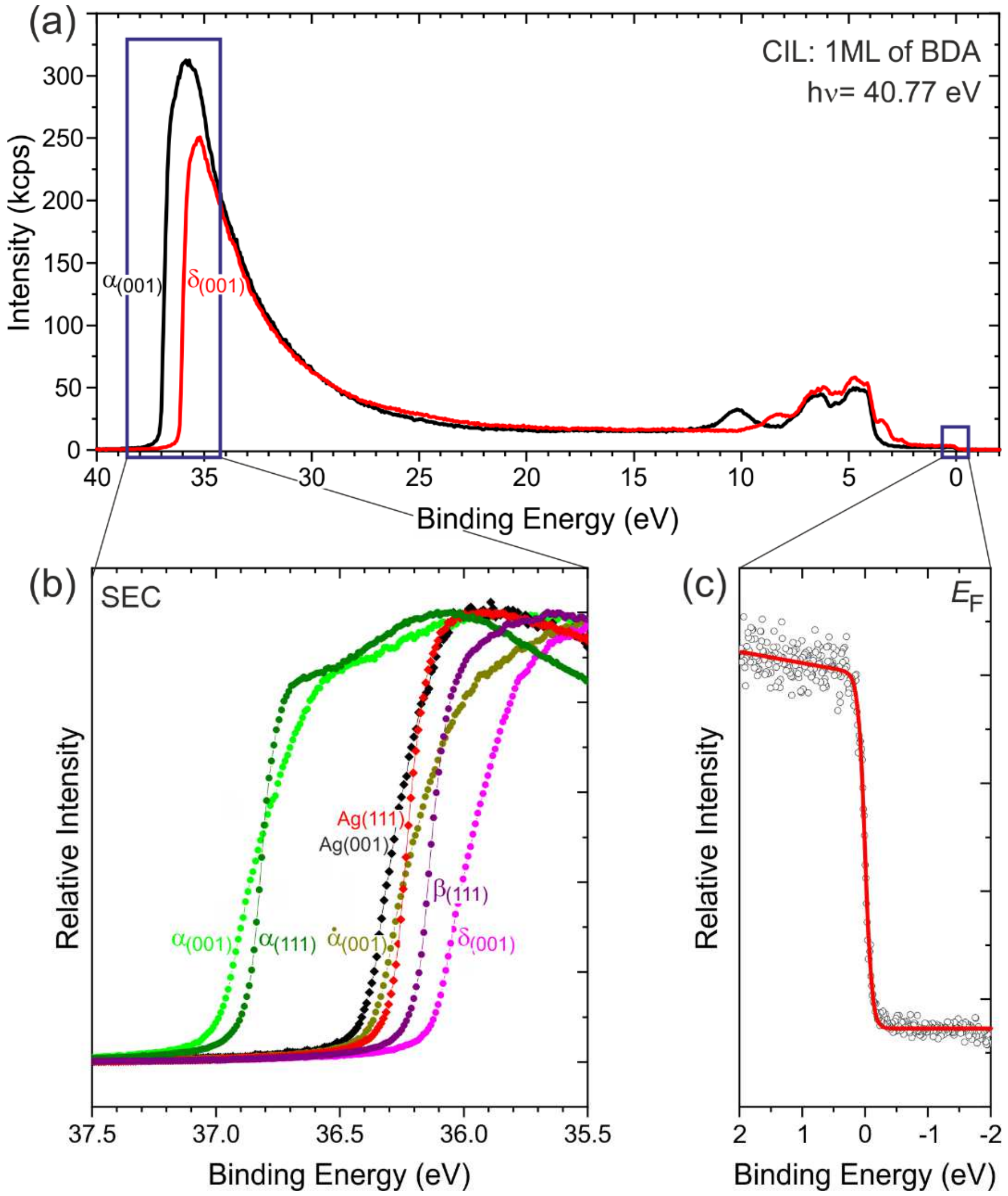

**Figure S6**: Work function measurement of CILs. (a) Overview of the valence band region measured with a 7 V sample bias (corrected to $E_{\mathrm{F}} = 0$ eV). (b, c) The details of SEC (b) and $E_{\mathrm{F}}$ (c) regions.

## 4. Mutual position of valence and O 1s peaks in the 2nd layer

Figure S7 shows that core-level binding energy shift is comparable to the shift of the peaks in the valence band region.

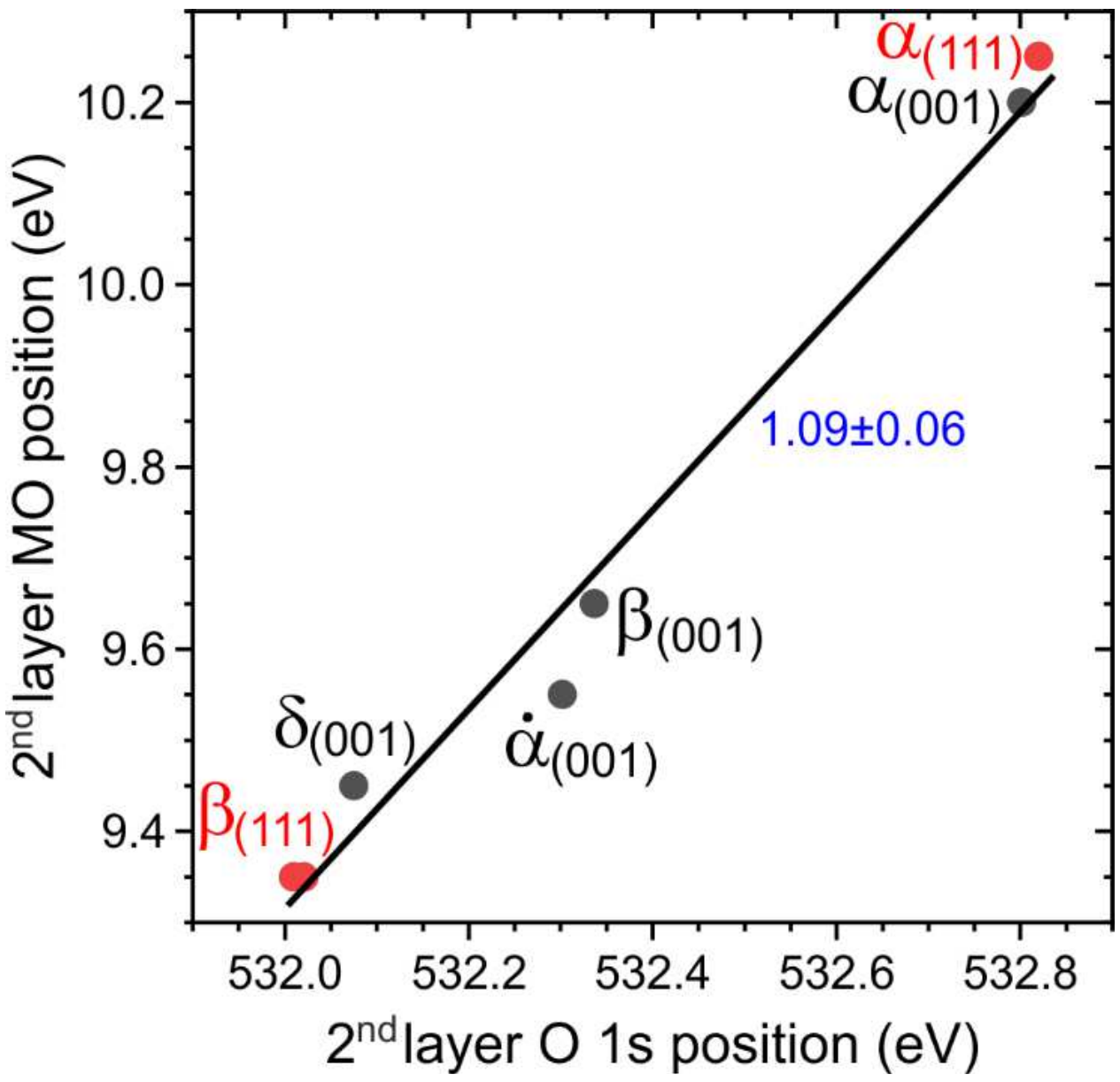

**Figure S7**: Position of 2nd layer peak associated with one of the highest lying occupied molecular orbitals (marked by a blue vertical line in Figures S3 and S5) on the position 2nd layer O2 component (Figures S2b and S4b) for all considered BDA CILs.

**5. Position of the overlayer peaks as a function of the overlayer thickness**

The evolution of the overlayer O 1s and C 1s peak positions with increasing BDA layer thickness is presented in Figure S8 for all considered CILs.

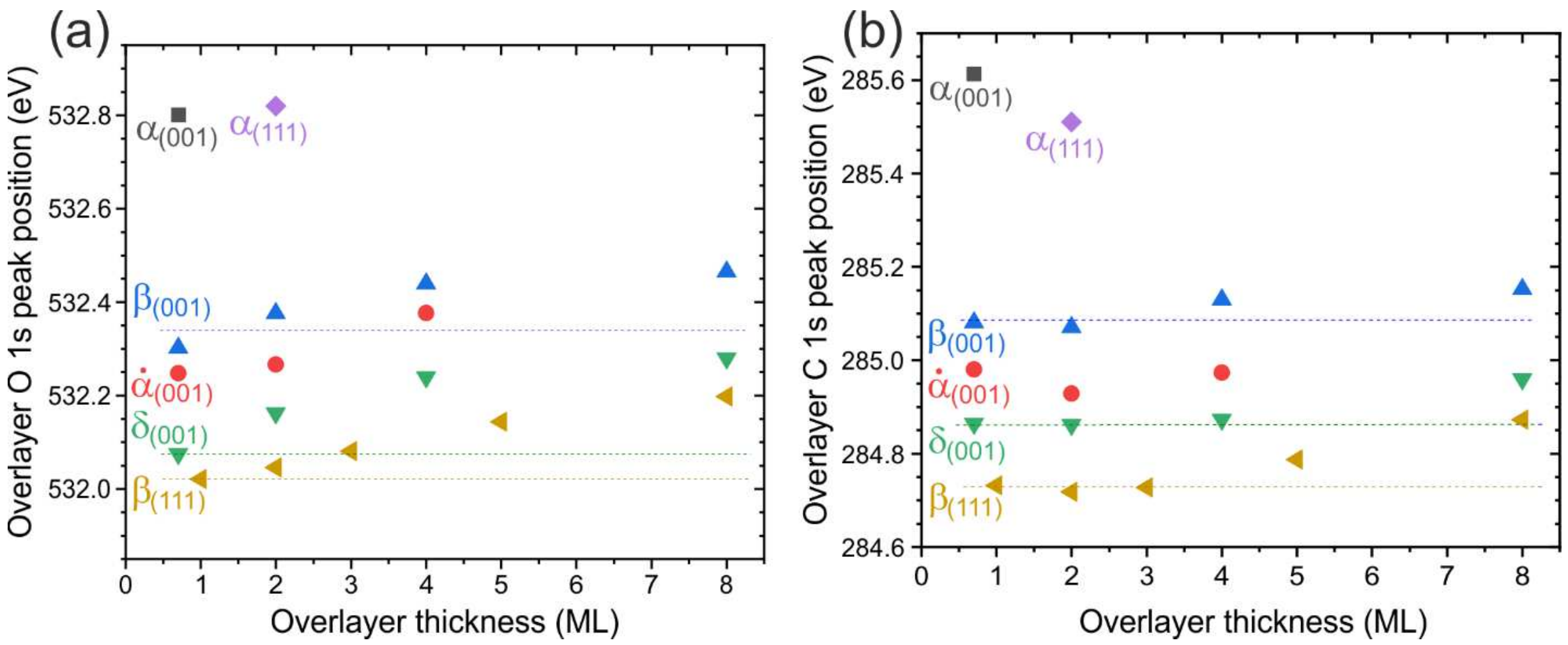

**Figure S8**: Position of (a) 2nd layer O 1s peak (O2 component, marked with a vertical line in Figure S2b) and (b) 2nd layer C 1s peak (C1 component, marked with a vertical line in Figure S1b) as a function of overlayer (i.e., 2nd layer and above) thicknesses deposited on BDA CILs.

## 6. References

[1] P. Procházka, M. A. Gosalvez, L. Kormoš, B. de la Torre, A. Gallardo, J. Alberdi-Rodriguez, T. Chutora, A. O. Makoveev, A. Shahsavar, A. Arnau, P. Jelínek, and J. Čechal, *Multiscale Analysis of Phase Transformations in Self-Assembled Layers of 4,4'-Biphenyl Dicarboxylic Acid on the Ag(001) Surface*, ACS Nano **14**, 7269 (2020).

[2] L. Kormoš, P. Procházka, A. O. Makoveev, and J. Čechal, *Complex K-Uniform Tilings by a Simple Bitopic Precursor Self-Assembled on Ag(001) Surface*, Nat. Commun. **11**, 1856 (2020).

[3] A. Schöll, Y. Zou, M. Jung, T. Schmidt, R. Fink, and E. Umbach, *Line Shapes and Satellites in High-Resolution x-Ray Photoelectron Spectra of Large π-Conjugated Organic Molecules*, J. Chem. Phys. **121**, 10260 (2004).

[4] M. Häming, A. Schöll, E. Umbach, and F. Reinert, *Adsorbate-Substrate Charge Transfer and Electron-Hole Correlation at Adsorbate/Metal Interfaces*, Phys. Rev. B **85**, 235132 (2012).

[5] A. Makoveev, P. Procházka, A. Shahsavar, L. Kormoš, T. Krajňák, V. Stará, and J. Čechal, *Kinetic Control of Self-Assembly Using a Low-Energy Electron Beam*, Appl. Surf. Sci. 154106 (2022).

[6] A. O. Makoveev, P. Procházka, M. Blatnik, L. Kormoš, T. Skála, and J. Čechal, *Role of Phase Stabilization and Surface Orientation in 4,4'-Biphenyl-Dicarboxylic Acid Self-Assembly and Transformation on Silver Substrates*, J. Phys. Chem. C **126**, 9989 (2022).